\documentclass[11pt]{article}
\usepackage[english]{babel}
\usepackage[utf8x]{inputenc}
\usepackage[T1]{fontenc}
\usepackage[bitstream-charter]{mathdesign}
\usepackage{amsmath}
\usepackage{graphicx}
\usepackage[left=2.5cm,right=2.5cm,top=2cm,bottom=2cm]{geometry}
\usepackage[usenames,dvipsnames]{xcolor}
\usepackage[colorlinks=true,linkcolor=Blue,linktoc=section,
citecolor=MidnightBlue,citebordercolor=MidnightBlue,
urlcolor=MidnightBlue,urlbordercolor=MidnightBlue]{hyperref}
\usepackage[small]{caption}
\usepackage[]{subfigure}
\usepackage{cite}
\usepackage{tikz}
\usetikzlibrary{decorations.pathmorphing,decorations.markings,arrows,patterns,shapes.misc}
\tikzset{
every picture/.style={line width=0.5pt},
photon/.style={decorate, draw=black,
  decoration={snake, amplitude=2pt, segment length=6pt}
},
fermion/.style={draw=black, postaction={decorate}, 
  decoration={markings, mark=at position .6 with
    {\arrow[scale=1]{triangle 45}}
  }
},
scalar/.style={draw=black},
cross/.style={cross out, draw=black, minimum
  size=2*(#1-\pgflinewidth), 
  inner sep=0pt, outer sep=0pt},
cross/.default={1pt}
}

\def\s0{\sigma_0}

\def\noi{\noindent}

\newcommand{\beqn}{\begin{eqnarray}}
\newcommand{\eeqn}{\end{eqnarray}}
\newcommand{\bea}{\begin{eqnarray}}
\newcommand{\ena}{\end{eqnarray}}

\newcommand{\neuti}{{\tilde{\chi}}_i^0}
\newcommand{\neutj}{{\tilde{\chi}}_j^0}
\newcommand{\neuto}{{\tilde{\chi}}_1^0}
\newcommand{\mneuto}{m_{{\tilde{\chi}}_1^0}}

\newcommand{\mchargopm}{m_{{\tilde{\chi}}_1^\pm}}

\newcommand{\mchargtpm}{m_{{\tilde{\chi}}_2^\pm}}
\def\mchargo{m_{\tilde{\chi}_1^+}}
\def\mchargt{m_{\tilde{\chi}_2^+}}

\def\nnzz{\neuto \neuto \to ZZ}
\def\nnww{\neuto \neuto \to W^+ W^-}
\def\nnff{\neuto \neuto \tof \bar f}
\def\SloopS{\texttt{SloopS}}

\def\nnff{\neuto \neuto \to f \bar f}

\newcommand{\tb}{t_\beta}

\begin{document}

\begin{titlepage}
\begin{center}

\vspace*{3cm}

{\Large {\bf Relic density calculations beyond tree-level, exact calculations versus effective couplings: the ZZ final state }}

\vspace{8mm}

{F. Boudjema${}^{1)}$, G. Drieu La Rochelle${}^{2)}$ and A. Mariano${}^{1)}$}\\

\vspace{4mm}

{\it 1) LAPTh$^\dagger$, Univ. de Savoie, CNRS, B.P.110,
F-74941 Annecy-le-Vieux , France}

{\it 2) Univ. de Lyon, F-69622 Lyon, France; Univ. Lyon 1, Villeurbanne\\CNRS/IN2P3, UMR5822, Institut de Physique Nucl\'eaire de Lyon}
\vspace{10mm}

\today
\end{center}

\centerline{ {\bf Abstract} } \baselineskip=14pt \noindent

{\small The inferred value of the relic density from cosmological
  observations has reached a precision that is akin to that of the LEP
  precision measurements. This level of precision calls for the
  evaluation of the annihilation cross sections of dark matter that
  goes beyond tree-level calculations as currently implemented in all
  codes for the computation of the relic density. In
  supersymmetry radiative corrections are known to be large and thus must be
    implemented. Full one-loop radiative corrections for many
  annihilation processes have been performed. It is important to
  investigate whether the bulk of these corrections can be
  parameterised through an improved Born approximation that can be
  implemented as a selection of form factors to a tree-level
  code. This paper is a second in a series that addresses this
  issue. After having provided these form factors for the annihilation
  of the neutralinos into fermions, which cover the case of a bino-like
  LSP (Lightest Supersymmetric Particle), we turn our attention here
  to a higgsino-like dark matter candidate through its annihilation into
  $ZZ$. We also investigate the cases of a mixed LSP. In all cases we
  compare the performance of the form factor approach with the result
  of a full one-loop correction. We also study the issue of the
  renormalisation scheme dependence. An illustration of the phenomenon
  of non decoupling of the heavy sfermions that takes place for
  the annihilation of the lightest neutralino into $ZZ$ is also
  presented.  }

\vspace*{\fill}

\vspace*{0.1cm} \rightline{LAPTh-019/14}\rightline{LYCEN 2014-03}

\vspace*{1cm}

$^\dagger${\small UMR 5108 du CNRS, associ\'ee  \`a l'Universit\'e
de Savoie.} \normalsize

\vspace*{2cm}

\end{titlepage}

\renewcommand{\topfraction}{0.85}
\renewcommand{\textfraction}{0.1}
\renewcommand{\floatpagefraction}{0.75}

\section{Introduction}
There is circumstantial evidence\cite{Jarosik:2010iu, planck_2013,Percival:2009xn}
from different astrophysical and cosmological observations for the
existence of Dark Matter (DM). In a particle physics context, the DM
candidate can only be part of a theory of New Physics that has, alas,
been elusive at the colliders so
far\cite{lhc_susy_limits_june2011,LHCTeVScales}. The next runs of the LHC will
perhaps tell us whether the Higgs with a mass of 125~GeV is part of a
richer underlying spectrum. Apart from its possible connection with
the Higgs, in particular bringing in a solution to the naturalness
problem, this New Physics may perhaps shed some light on the nature of
Dark Matter.  Although current LHC data set the scale of many New
Physics scenarios in the TeV range\cite{LHCTeVScales}, this inferred
large scale refers in fact to the non observation of new coloured
particles. On the one hand, the mass of the non coloured weakly interacting dark matter
candidate is much less constrained from LHC analyses. On the other hand measurements of
the relic density are now accurate at the per-cent
level\cite{planck_2013} and provide very strong constraints on the
properties of DM. This supposes of course that we know the
thermodynamics and cosmology of the universe, the standard approach
for example incorporates thermal production but there are
alternatives to the standard approach\cite{nonconventional-relic}. In any case, considering the
per-cent precision on the relic density measurement, one needs to
provide, from the particle physics side, dark matter annihilation cross
sections at the per-cent level or better. State of the art
codes\cite{micromegas,darksusy,superiso-relic,boudjema_gondolo} for
the calculation of the relic density have
been developed during the last decade. They are practically all based
on tree-level cross sections and are therefore not precise enough. In
some instances large corrections to the DM annihilation cross sections
occur, for instance the classical
Sommerfeld\cite{Hisano:2002fk,Iengo:2009ni,ArkaniHamed:2008qn,Hryczuk:2011vi,Sommerfeldinclude}
effect or the electroweak Sudakov effect with the concomitant
inclusion of final state
radiation\cite{boudjema_chalons1,Ciafaloni:2013hya}. These special
effects are common to cases with TeV and above DM due to the presence
of two disparate scales, the DM mass and a low mass mediator for which
an example is the $W$ gauge boson. In these regimes, the leading
corrections that take into account these effects can be extracted. It
remains that there are also important corrections in much more general
situations irrespective of the mass of the DM. These corrections are
far from being negligible, as shown in
\cite{baro07,baro09,boudjema_chalons1,Freitas-relic-qcd,HerrmannQCD}. \\

For the last few years some of us have set up a
programme\cite{Boudjema:2005hb,baro07,boudjema_chalons1} for
the calculation of the full one-loop electroweak corrections for
practically all annihilation channels of the Dark Matter candidate in
the minimal supersymmetric model (MSSM). The first important
ingredient of this programme requires a coherent and flexible
renormalisation of
all sectors of the MSSM, allowing for different renormalisation schemes. To be able to handle the large variety of
possible annihilation and co-annihilation channels of the neutralino, 
a tool for the automated calculation of one-loop corrections in the MSSM was developed.
This tool, \texttt{SloopS}~\cite{Boudjema:2005hb,Sloops-higgspaper,baro09,boudjema_chalons1}, 
based on \cite{lanhep,feynarts,formcalc,looptools}, allows to perform full one-loop
calculations \cite{baro07,Sloops-higgspaper,baro09,boudjema_chalons1}. 
Ultimately the aim is to implement these corrections in a code such
as {\tt micrOMEGAs}~\cite{micromegas} thus improving on the tree-level
calculation of the annihilation cross sections.  The difficulty is
that one needs to correct some 3000 processes at one-loop in the case
of the MSSM. Considering the large number of fields and parameters,
the one-loop correction for each process requires computing a few
hundred to a few thousands Feynman diagrams at one-loop. This is  far
more demanding that a computation at leading order. This is also totally
intractable, for instance, in a scan over the parameter space. Yet, one can inquire
whether the bulk of these corrections could be captured in a more
compact form through effective form factor corrections to tree-level
couplings. This would mean that these corrections are universal in the
sense of being process independent and amount to an overall shift of
the couplings of the different tree-level vertices. This improved
Born approximation works quite well for LEP observables. Part of these
corrections are for example due to the running of coupling constants
but there might be other genuine corrections. If such an approximation
works one could very easily recycle these improved Born couplings for
any process which would very much improve a code such as {\tt
  micrOMEGAs}. In Ref.~\cite{Boudjema:2011ig} we explored whether this
approach works in the case of the most simple of all annihilation
channels, $\nnff$, the annihilation of the neutralino into fermion
pairs. In that case we introduced the effective vertices $\neuto f
\tilde{f}$ and $\neuto \neuto Z$ together with the improved vertex
$Z f \bar f$. Some of these effective vertices had also been discussed
in~\cite{Hollik_susyeff,nojiri_nondecoupling} outside the context of
dark matter annihilation and later in \cite{Akcay:2012db}. Our first
study revealed some very interesting results. In the case of a
bino-like neutralino the percentage correction in the effective approach turned out to be a very
good approximation falling short of about $2\%$ compared to
the full one-loop calculation. However as the bino component drops, the
effective approach is a very rough approximation that worsens as the
mass of the LSP increases, this is due to "non universal" rescattering
effects (for example, box diagrams obtained from $\neuto \neuto \to W^+ W^-$
followed by $ W^+ W^- \to f \bar f$). However in these cases, the
important point is that as the bino component drops the $\nnff$ is not
an efficient annihilation channel, the largest channels being by far 
the annihilation processes of the
neutralino into vector boson pairs.There is  
typically a factor of $10^4$ between the two cross sections. In these scenarios, it is much
more important to concentrate on annihilation to vector bosons, take
the radiative corrections to $\neuto \neuto \to W^+ W^-, ZZ$ into
account and implement $\neuto \neuto \to f \bar f$ at tree-level, if at
all. We must therefore pursue the implementation of an effective couplings
approach to the annihilation into vector bosons. The aim of the present
analysis is to concentrate first on $\neuto \neuto \to ZZ$ before
presenting our results for $\neuto \neuto \to W^+ W^-$, this will help
us identify some new features without the need to worry, for example,
about QED corrections and other complications that usually affect
processes with charged particles. Another important point is that radiative
corrections to neutralino annihilation are sensitive, even though only
logarithmically, to the presence of heavy sfermions with masses far
above those of the LSP. We  discussed this point when we  computed
the radiative corrections to $\nnff$. So even if no coloured particles
have been discovered at the LHC, meaning they may be in the TeV range,
neutralino annihilation,
if measured precisely, does probe their effect. \\

As we will see, the process $\neuto \neuto \to ZZ$ is relevant when the $\neuto$ has a
fair amount of higgsino component. Therefore in our applications we
consider a neutralino LSP with a mass in excess of 110~GeV to conform with
the LEP limit on the chargino mass. For bino-like LSP this limit does not
apply but, as we will see, for a bino-like LSP the process $\nnzz$ is not relevant. The
LHC\cite{CMS:2013dea,TheATLAScollaboration:2013hha,ATLAS:2013rla,CMS:2013bda}
may also provide some more stringent limits on the
chargino/neutralino, however these are often model dependent. Even
when LHC analyses are recast within a simplified
model\cite{Kraml:2013mwa,Aad:2014vma}, the limits are not necessarily applicable to our
set-up, for example we work here with very heavy sleptons and
squarks. Moreover the higgsino scenario is extremely challenging because of the small energy 
that is left for the visible tracks due to the small mass splitting within the higgsino system.\\

We will write $\tb$ for $\tan \beta$. $\tb$, at tree-level, measures
the ratio of the vacuum expectation values in the up to down
sector. The crucial parameters for calculating $\nnzz$ are those of
the neutralino/chargino sector, namely $M_1,M_2$ (respectively the
$U(1)$ and $SU(2)$ soft gaugino masses at the electroweak scale) and
$\mu$, the higgsino mass parameter. Apart from these 3 parameters, and
unless explicitly stated, we fix all other SUSY soft masses. The mass
of the $SU(3)$ gaugino, $M_3$, is set to 1~TeV.  We take $\tb=10$ and a
common sfermion mass for both the left and right sfermions,
$m_{\tilde{f}}=800$~GeV. The tri-linear mixing parameter is set at
$A_f=2$~TeV for all sfermions. This high value is in fact of relevance
for the third generation squarks, in particular the stop. The reason
we take this value is to reproduce a Higgs mass in accord with that of
the observed Higgs at the LHC. With our default parameters we obtain
$M_h=121.5$~GeV. This could be easily increased by taking heavier stops
as will be done when we will study the ``non decoupling''
universal effects. Moreover, as we will see,  Higgs exchange is subdominant in the scenarios we
will cover and therefore, for the cross sections we study,  sensitivity to the Higgs masses is negligible. \\
All cross sections are calculated for a centre of mass energy $s = 4
M_{\neuto}^{\;\;2}/(1-v^2)$ with $v=0.1$, which gives a relative
scattering velocity $v_{{\rm rel}} =0.198\sim 0.2$ as befits a relic density calculation. $M_{\neuto}$ is
the mass of the neutralino LSP.

\section{$\nnzz$ at tree-level, renormalisation and effective vertices}
\subsection{$\nnzz$ at tree-level}
\label{subsec:tree-level}
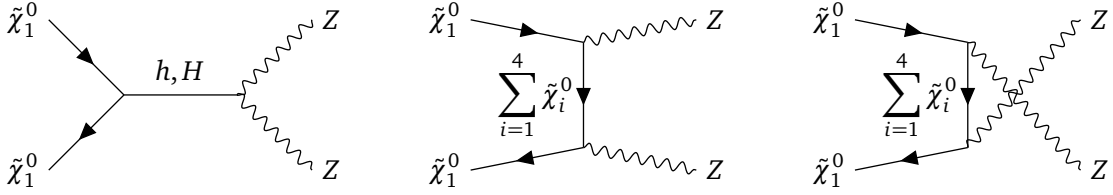
\begin{figure}[h]
\centering
\begin{tikzpicture}[scale=1.0]
\draw[fermion] (-1,1) -- node[label={[at start]180:$\neuto$}] {} (0,0);
\draw[fermion] (0,0) -- node[label={[at end]-180:$\neuto$}] {} (-1,-1);
\draw[scalar] (0,0) -- node[label={above:$h,H$}] {} (1.5,0);
\draw[photon] (2.5,1) -- node[label={[at start]0:$Z$}] {} (1.5,0);
\draw[photon] (1.5,0) -- node[label={[at end]0:$Z$}] {} (2.5,-1);
\end{tikzpicture}
\hspace{.7cm}
\begin{tikzpicture}[scale=1.0]
\draw[fermion] (-1,1) -- node[label={[at start]180:$\neuto$}] {} (0.5,0.7);
\draw[fermion] (0.5,0.7) -- node[label={180:$\displaystyle\sum_{i=1}^{4} \neuti$}] {} (0.5,-0.7);
\draw[fermion] (0.5,-0.7) -- node[label={[at end]-180:$\neuto$}] {} (-1,-1);
\draw[photon] (0.5, 0.7) -- node[label={[at end]0:$Z$}] {} (2,  1);
\draw[photon] (0.5,-0.7) -- node[label={[at end]0:$Z$}] {} (2,-1);
\end{tikzpicture}
\hspace{.7cm}
\begin{tikzpicture}[scale=1.0]
\draw[fermion] (-1,1) -- node[label={[at start]180:$\neuto$}] {} (0.5,0.7);
\draw[fermion] (0.5,0.7) --  node[label={180:$\displaystyle\sum_{i=1}^{4}\neuti$}] {} (0.5,-0.7);
\draw[fermion] (0.5,-0.7) -- node[label={[at end]-180:$\neuto$}] {} (-1,-1);
\draw[photon] (0.5, 0.7) --  node[label={[at end]0:$Z$}] {} (2,-1);
\draw[photon] (0.5,-0.7) --  node[label={[at end]0:$Z$}] {} (2,  1);
\end{tikzpicture}
\caption{{\em Diagrams contributing to the $\neuto \neuto \to ZZ$
  annihilation at the tree level. $h,H$ denote the lightest and
  heaviest CP-even Higgs, respectively.}}
\label{fig:tree-diagrams}
\end{figure}
The diagrams contributing to $\nnzz$ at tree-level are shown in
Fig.~\ref{fig:tree-diagrams}. In practically all cases, and certainly when $\nnzz$ is
a relevant annihilation channel, the largest contribution proceeds
through the $t$-channel diagrams via the exchange of all
neutralinos. The Higgs exchange contribution is then subdominant. The cross
section is therefore largely driven by the strength of the
$\neuto\neuti Z\;\;( i=1,..4$) vertex.
First of all, at tree-level one has only the following structure
\beqn
{\cal {L}}_{\neuti \neutj Z}^0=\frac{g_Z}{4} \neuti
\Big\{\Big(N_{i3}N_{j3}^\star-N_{i4} N_{j4}^\star \Big)  \gamma_\mu P_R\;-\;
\Big(N_{i3}^\star N_{j3}-N_{i4}^\star N_{j4} \Big)  \gamma_\mu P_L \Big\} \neutj Z^\mu, \quad \quad g_Z=\frac{e}{c_W s_W},
\label{eq:nnz-tree}
\eeqn
where $P_{R/L}=1/2 (1\pm \gamma_5)$. $N$ is the unitary complex
matrix that defines the physical fields $\tilde\chi^0_i, (i=1,\dots,4)$ in
terms of the interaction eigenstates
$\big(\psi^{n}\big)^t=(\tilde{B}^0,\tilde{W}^{0},
\tilde{H}_1^0,\tilde{H}_2^0)$ (respectively bino, wino and higgsinos)
\begin{eqnarray}
\label{eq:N-neut} \tilde\chi^{0}=N \psi^{n}.
\end{eqnarray}
$N$ diagonalises the mass mixing matrix of the neutralino
sector $Y$, see~\cite{baro09} for details and conventions. \\
As can be seen from Eq.~\ref{eq:nnz-tree}, the strength of the $\neuti
\neutj Z$ coupling is solely related to the higgsino component (note
the presence of the matrix elements $N_{i3}$ and $N_{i4}$ in it). Therefore $\nnzz$ is expected
to play an important role, as an efficient annihilation channel, for a higgsino-like neutralino.  From
Eq.~\ref{eq:N-neut} we define
the amount of bino, wino and higgsino of $\neuto$ as follows 
\beqn 
\tilde\chi^0_1= N_{11} \tilde{B}^0 + N_{12} \tilde{W}^0 + N_{13} \tilde{H}^0_1 + N_{14} \tilde{H}^0_2,
\eeqn
which defines the bino, wino and higgsino fraction of $\neuto$ as
$f_B=|N_{11}|^2, f_W=|N_{12}|^2, f_H=|N_{13}|^2+|N_{14}|^2$ respectively. The
composition of the other neutralinos is defined in an analogous way.

\begin{figure}[tbp]
\centering
higgsino-like 1, $M_1=500$~GeV, $M_2=1000$~GeV\\
\includegraphics{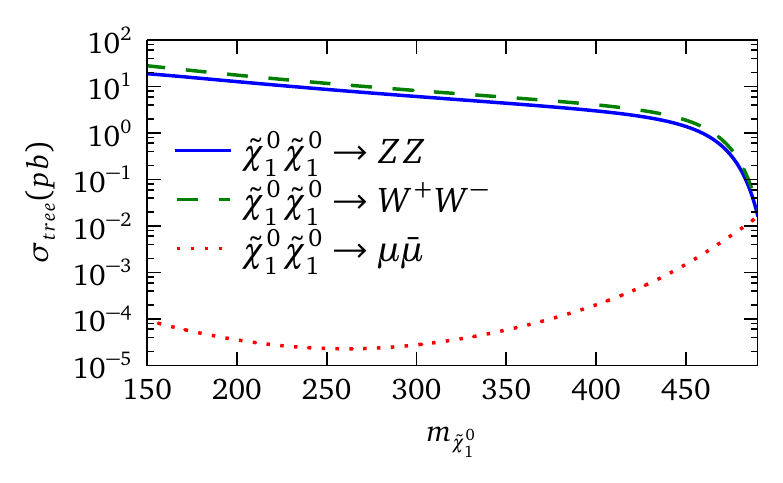}
\includegraphics{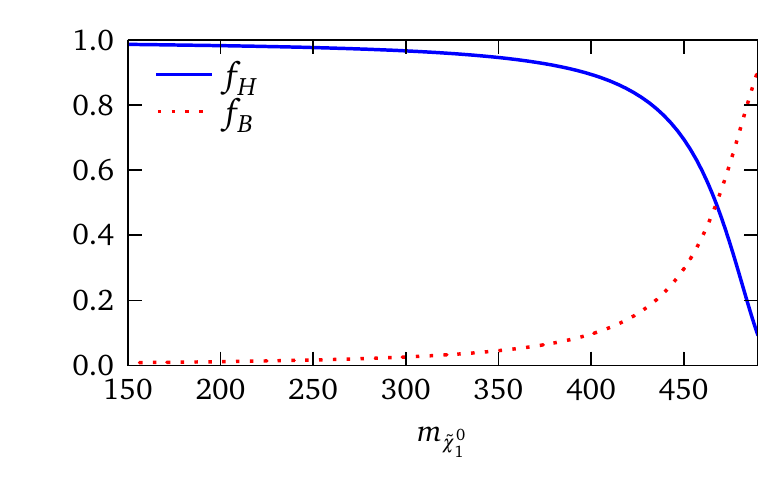}\vspace{-0.3cm}\\
higgsino-like 2, $M_2=500$~GeV, $M_1=1000$~GeV\\
\includegraphics{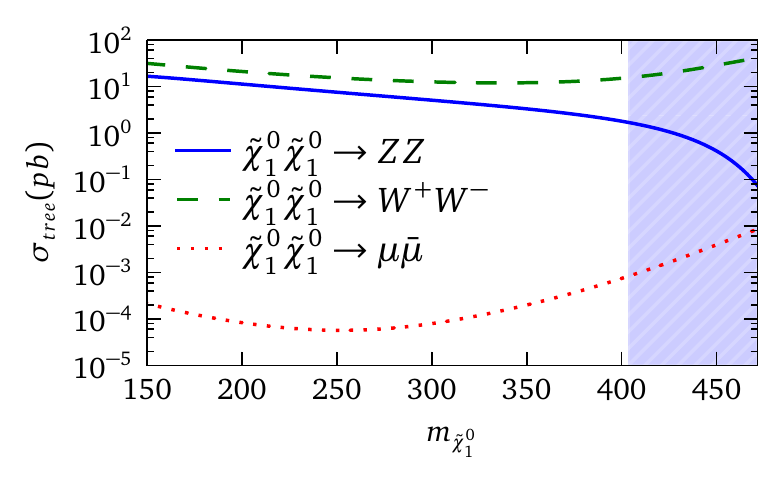}
\includegraphics{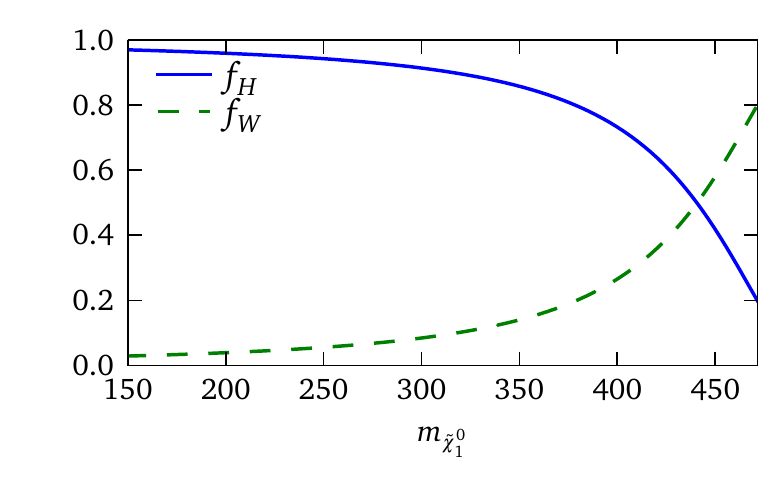}\vspace{-0.3cm}\\
bino-like, $M_2=500$~GeV, $\mu=1000$~GeV \\
\includegraphics{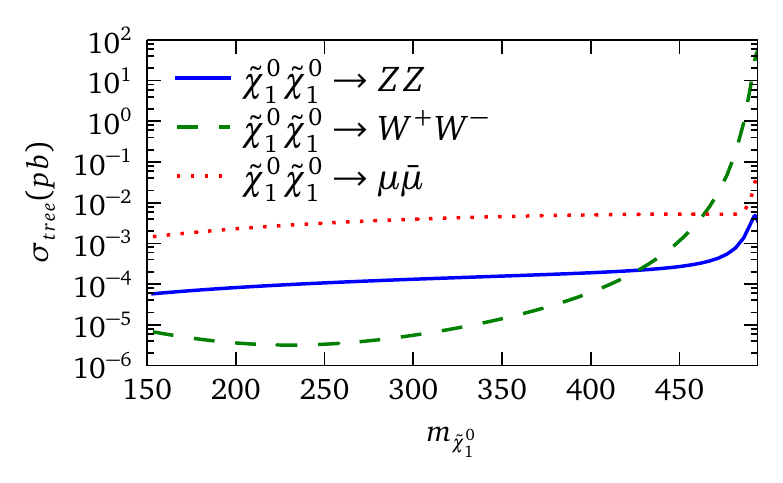}
\includegraphics{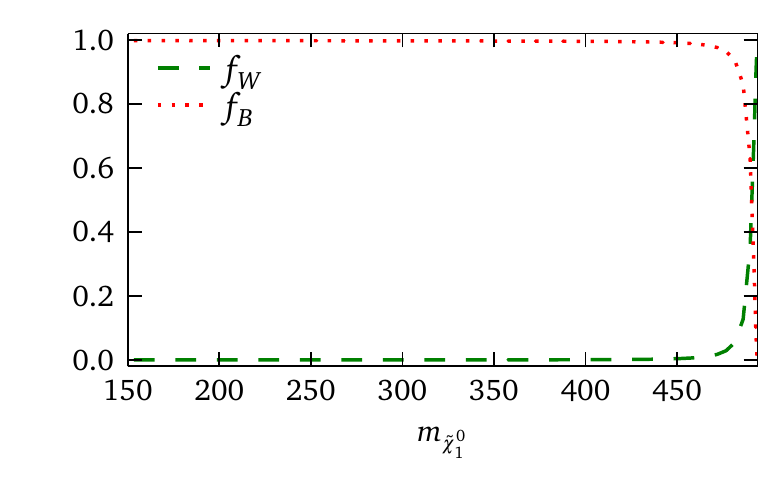}\vspace{-0.3cm}\\
wino-like, $M_1=500$~GeV, $\mu=1000$~GeV \\
\includegraphics{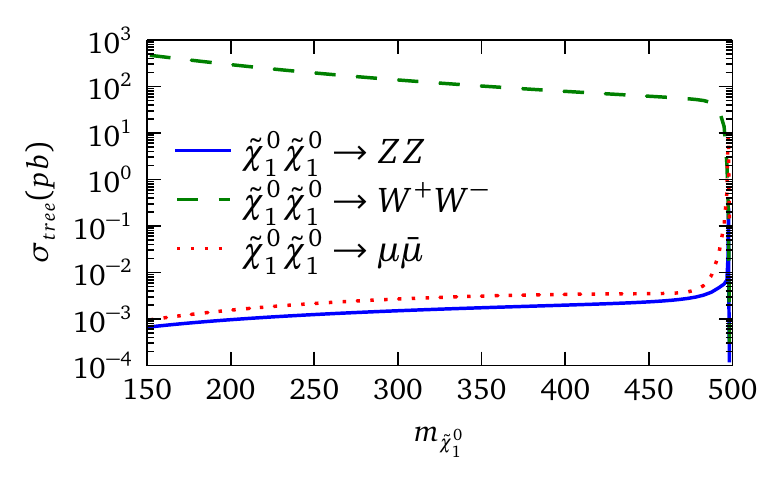}
\includegraphics{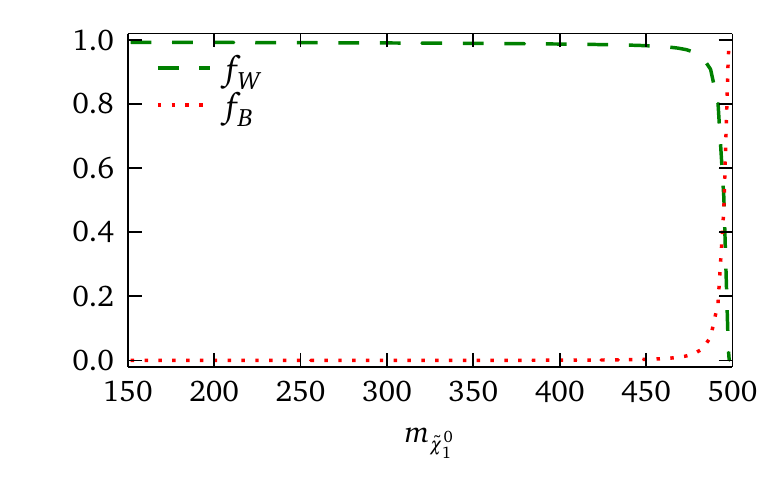}
\caption{{\em Tree-level annihilation cross sections (left panels) for
    a relative velocity of $v\simeq 0.2$ (see text) of the neutralino
    LSP for different masses and compositions (right panel).
    The characterisation higgsino-like, bino-like, wino-like stands for
    masses of the
    LSP below 400~GeV or so.  The blue shaded area in the first
    figure of the second row corresponds to a contribution of $\nnzz$
    less than $10\%$ of the total annihilation cross section (see
    text). For the bino and wino cases (3rd and 4th row) this is
    always the case. For the higgsino-like 1, first row, annihilation
    into $ZZ$ is always larger than $10\%$ in the mass range shown.}}
\label{fig:tree-level}
\end{figure}
We consider two cases of higgsino dominance. The higgsino-like
scenario 1 is obtained by having $M_1=500$~GeV and $M_2=1$~TeV while $\mu$
is varied from 150 to 600~GeV. In this variation the LSP composition
goes from higgsino-like to bino-like. As can be seen from
Fig.~\ref{fig:tree-level} the change of the nature of the LSP is quite
sudden and occurs at an LSP mass around 450~GeV. \\
In the second scenario, the higgsino-like scenario 2, we swap the values
of $M_1$ and $M_2$ as compared to the first scenario, thus allowing
for a transition from higgsino to wino, which occurs in this case
around $400$~GeV.\\
The last scenarios are bino-like and wino-like with $\mu=1$~TeV while
$M_1$ and $M_2$ are much smaller.  \\
Fig.~\ref{fig:tree-level} confirms
that $\nnzz$ is indeed important for a higgsino-like LSP, being of
order $10^2$pb for $f_H \sim 1$. In this case the $\nnzz$ cross
sections is of the same order as that of $\nnww$ and is as much as 6
orders of magnitude larger than the cross section for the annihilation into fermion pairs. This
overwhelming dominance is still present even when the higgsino content
drops to $50\%$. At this level of higgsino content, $\nnzz$ drops fairly quickly. In
the first scenario, as some bino contamination feeds in for the highest LSP masses, $\nnff$ starts 
picking up (due to the right slepton contribution) but without
becoming competitive with $\nnzz$ which remains of the same order as $\nnww$. Note that in some scenarios and for some
masses there might be other processes that enter the relic density
calculation that are more effective, we can think of co-annihilation
processes for example. We do not show them here. It is true that if
these were non negligible they would reduce the weight of the
$\nnzz$. Each of the cross sections we display tracks a particular
composition of the LSP: $\nnff$ the bino, $\nnzz$ the higgsino and
$\nnww$ the wino (even if the latter also contributes
substantially in the higgsino case). Note that when we talk about $\nnff$ we have essentially in mind 
the annihilation into charged leptons through right slepton exchange because  they have the largest hypercharge.\\
The situation is somehow different in the second scenario: here, as expected,
$\nnww$ shoots up as the wino content increases while $\nnzz$ drops
continuously by 2 orders of magnitude within a range of 150~GeV in
LSP mass and $\nnff$ is always negligible. The wino nature starts affecting
the cross-section at $f_H=0.8$. This is easy to understand. In the
pure wino limit $\nnww$ is by far overwhelming, see for example the
last set of figures in Fig.~\ref{fig:tree-level} for the wino-like
scenario ($M_2 \ll (M_1,\mu)$). In the bino dominated region (scenario
3, $M_1 \ll (M_2,\mu)$) , the largest cross section is $\nnff$,
$\nnzz$ being a fraction of it, however as soon as there
is even a small amount of wino component, $\nnww$ takes off. In the
wino dominated region (scenario 4, $M_2 \ll (M_1,\mu)$), $\nnww$ is
the only cross section of relevance being 6 orders of magnitude
larger than $\nnzz$. Here it is pointless to consider radiative
corrections to $\nnzz$ for the purpose of improving the calculation of
the relic density.

As an aside, let us mention that the smallness of $\nnff$ cross
section is not only due to how small the bino content is in the
higgsino region. Even when the bino content is large, the cross
section is small compared to what we see for $\nnzz$ in the higgsino
region. Couplings left aside, compared to $\nnzz$ and $\nnww$, $\nnff$ for
massless fermions suffers from a chiral suppression that leads to a
vanishing $s$-wave contribution. Moreover, $t$-channel processes are larger as the
spin of the exchanged $t$-channel particle is large. The latter
point gives an advantage to $\nnzz$ mediated by a fermion rather than
$\nnff$ mediated by a sfermion, not to mention the fact that the
exchanged sfermion is generally much heavier than the $\neuto$ whereas
for $\nnzz, \nnww$ in the higgsino region there is at least one of the
neutralinos/charginos of very comparable mass to the LSP. Let us point
also at another feature. In the higgsino region, say $\mu=200$~GeV with
$M_1=M_2/2=500$~GeV,  a $20\%$ change in $M_1$ results in about $3\%$
change in the cross section while the neutralino mass hardly
changes. In the mixed region with $\mu=450$~GeV, $M_1=M_2/2=500$~GeV, a
$20\%$ change in $M_1$ results in practically $100\%$ change in the cross
section, while the mass of the LSP becomes as much as 50~GeV smaller. This means that
the determination of $M_1$ (and $\mu$) is crucial in this region. This
will have a consequence on the scheme dependence in the neutralino
sector. The $\tb$ dependence is extremely mild either in the pure or
mixed regions. From the observations we have just made on the
tree-level cross sections, it is to be stressed that it is for
higgsino-like LSP configurations that $\nnzz$ is of relevance and it
is for these configurations that the effective approach we are seeking
should best approximate the full one-loop corrections.

Let us note that we have shown all cross sections as a function of the
LSP mass, that is as a function of physical parameters that could be
measured instead of the underlying parameters $M_1,M_2,\mu$. The
nature of the LSP is given by $f_H,f_W$, the higgsino and wino
content. The latter could in principle be reconstructed from the decay
of other neutralinos or the production of neutralinos at colliders. This is also important 
when we move to implementing the radiative
corrections, where the physical masses, in particular that of the LSP,
will be used as input parameters. 

\subsection{Beyond the tree-level, full {\it vs} the Form Factor effective approach}
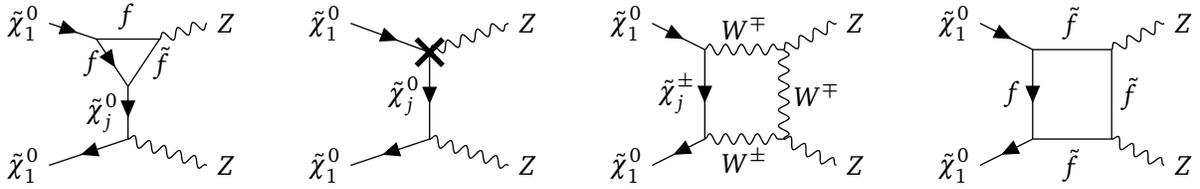
\begin{figure}[h]
\centering
\begin{tikzpicture}[scale=0.7]
\draw[fermion] (-1.5,1) -- node[label={[at
  start]180:$\tilde\chi^0_1$}] {} (-0.6,0.7);
\draw[fermion] (0,-0.2) -- node[label={180:$\tilde\chi^0_j$}] {} (0,-1.2);
\draw[fermion] (0,-1.2) -- node[label={[at end]180:$\tilde\chi^0_1$}]
{} (-1.5,-1.7);
\draw[fermion] (-0.6,0.7) -- node[label={180:$f$}] {}(0,-0.2);

\draw[scalar] (-0.6,0.7) -- (0.6,0.7) node[label={[midway]90:$\tilde
  f$}] {} -- (0,-0.2) node[label={[midway]0:$\tilde f$}] {};
\draw[photon] (0.6,0.7) -- node[label={[at end]0:$Z$}] {} (1.5,1);
\draw[photon] (0,-1.2) -- node[label={[at end]0:$Z$}] {} (1.5,-1.7);
\end{tikzpicture}
\hspace{0.5cm}
\begin{tikzpicture}[scale=0.7,cross/.style={cross out, draw=black, minimum
    size=2*(#1-\pgflinewidth), inner sep=0pt, outer sep=0pt, line width=2pt},
  cross/.default={1pt}]

\draw[fermion] (-1.5,1) -- node[label={[at
  start]180:$\tilde\chi^0_1$}] {} (0,0.45);
\draw[fermion] (0,0.45) -- node[label={180:$\tilde\chi^0_j$}] {} (0,-1.2);
\draw[fermion] (0,-1.2) -- node[label={[at end]180:$\tilde\chi^0_1$}]
{} (-1.5,-1.7);
\draw (0,0.45) node[cross=7pt]{}; 
\draw[photon] (0,0.45) -- node[label={[at end]0:$Z$}] {} (1.5,1);
\draw[photon] (0,-1.2) -- node[label={[at end]0:$Z$}] {} (1.5,-1.7);
\end{tikzpicture}
\hspace{0.5cm}
\begin{tikzpicture}[scale=0.7]
\draw[fermion] (-1.5,1) -- node[label={[at
  start]180:$\tilde\chi^0_1$}] {} (-0.5,0.5);
\draw[fermion] (-0.5,-1.2) -- node[label={[at
  end]180:$\tilde\chi^0_1$}] {} (-1.5,-1.7);
\draw[fermion] (-0.5,0.5) -- node[label={180:$\tilde\chi^\pm_j$}] {} (-0.5,-1.2);
\draw[photon] (-0.5,-1.2) -- node[label={-90:$W^\pm$}] {} (1,-1.2);
\draw[photon] (-0.5,0.5) -- node[label={90:$W^\mp$}] {} (1,0.5);
\draw[photon] (1,0.5) -- node[label={0:$W^\mp$}] {} (1,-1.2);
\draw[photon] (1,0.5) -- node[label={[at end]0:$Z$}] {} (2,1);
\draw[photon] (1,-1.2) -- node[label={[at end]0:$Z$}] {} (2,-1.7);
\end{tikzpicture}
\hspace{0.5cm}
\begin{tikzpicture}[scale=0.7]
\draw[fermion] (-1.5,1) -- node[label={[at
  start]180:$\tilde\chi^0_1$}] {} (-0.5,0.5);
\draw[fermion] (-0.5,-1.2) -- node[label={[at
  end]180:$\tilde\chi^0_1$}] {} (-1.5,-1.7);
\draw[fermion] (-0.5,0.5) -- node[label={180:$f$}] {} (-0.5,-1.2);
\draw[] (-0.5,-1.2) -- node[label={-90:$\tilde f$}] {} (1,-1.2);
\draw[] (-0.5,0.5) -- node[label={90:$\tilde f$}] {} (1,0.5);
\draw[] (1,0.5) -- node[label={0:$\tilde f$}] {} (1,-1.2);
\draw[photon] (1,0.5) -- node[label={[at end]0:$Z$}] {} (2,1);
\draw[photon] (1,-1.2) -- node[label={[at end]0:$Z$}] {} (2,-1.7);
\end{tikzpicture}
\caption{{\em A selection of one-loop diagrams that contribute to
    $\nnzz$. The first diagram is an example of a triangle of
    fermion/sfermion loops. The second is the implementation of a
    counterterm. The third is a box diagram of charginos and weak
    boson, note that is can be seen as $\nnww$ followed by $ W^+ W^-
    \to ZZ$. The last set consists of box diagrams of fermion/sfermions. }}
\label{fig:1loop-diagrams}
\end{figure}

$\nnzz$ at tree-level requires the computation of not more than 2 sets
of diagrams (see Fig.~\ref{fig:tree-diagrams}). Moreover when $\nnzz$
is an efficient annihilation channel, Higgs exchange is not
relevant. A one-loop computation of the same process calls for
hundreds of diagrams, a selection of these is shown in
Fig.~\ref{fig:1loop-diagrams}.

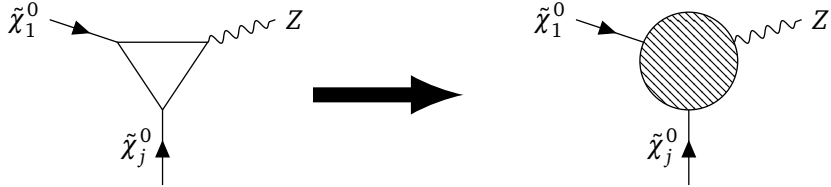
\begin{figure}[h]
\centering
\begin{tikzpicture}[scale=1.0]
\draw[fermion] (-1.5,1) -- node[label={[at
  start]180:$\tilde\chi^0_1$}] {} (-0.6,0.7);
\draw[fermion] (0,-1.2) -- node[label={180:$\tilde\chi^0_j$}] {} (0,-0.2);
\draw[scalar] (-0.6,0.7) -- (0.6,0.7) -- (0,-0.2) -- cycle;
\draw[photon] (0.6,0.7) -- node[label={[at end]0:$Z$}] {} (1.5,1);
\tikzstyle{fleche}=[->,>=latex,line width=2mm]
\draw[fleche] (2.0,0) -- (4,0);
\end{tikzpicture}
\hspace{0.5cm}
\begin{tikzpicture}[scale=1.0]
\draw[fermion] (-1.5,1) -- node[label={[at
  start]180:$\tilde\chi^0_1$}] {} (-0.6,0.7);
\draw[fermion] (0,-1.2) -- node[label={180:$\tilde\chi^0_j$}] {} (0,-0.2);
\draw[pattern=north west lines] (0,0.45) circle (0.65cm);
\draw[photon] (0.6,0.7) -- node[label={[at end]0:$Z$}] {} (1.5,1);
\end{tikzpicture}
\caption{{\em Example of the one-loop vertex diagrams that can be cast into an effective $\neuto \tilde\chi^0_j Z$ vertex.}}
\label{fig:1loop-diagrams-eff}
\end{figure}

The first two diagrams of Fig.~\ref{fig:1loop-diagrams} can be
considered as a universal correction to the $\neuto \tilde\chi^0_i Z$ vertex
through a fermion/sfermion loop to which counterterms are added. This is genuinely
universal since these fermion/sfermion states do not relate to the
external states.
Needless to say that box diagrams are most time consuming in a numerical evaluation. Therefore
if one can show that their contribution is small after all, the whole
one-loop correction could be cast into a correction to
vertices that were already needed for the tree-level
calculation.  If this is the case, one needs to replace the tree-level
vertices, in particular the most important ones, namely $\neuto
\tilde\chi^0_i Z$, by an effective form factor vertex as shown in
Fig.~\ref{fig:1loop-diagrams-eff}. \\

In this form factor approach, a single improved coupling could be used
for any process, thus allowing an easy adaptation of tree-level
codes.
Nonetheless this does assume that the effective $\tilde\chi^0_i \tilde\chi^0_i Z$
vertex is just a rescaled version of the tree-level vertex defined in Eq.~\ref{eq:nnz-tree}, with an overall
replacement ${\cal {L}}_{\neuti \neutj Z}^0 \rightarrow \kappa_{ij} \;
{\cal {L}}_{\neuti \neutj Z}^0 $, $\kappa_{ij}$ is the
form-factor. This form factor can be obtained, as we show later,
through modifications to the universal quantities $g_Z$ and a new
effective ``mixing matrix'' $N$ that replace the tree-level values of
Eq.~\ref{eq:nnz-tree} with $g_Z \to g_Z^{{\rm eff}}$ and $ N \to
N^{{\rm eff}}$. The approach does not allow for new Lorentz
structures, otherwise the concept of an improved Born approximation
would be meaningless.

At one-loop a full calculation of the one-particle irreducible (1PI)
$\neuti (k_i) \neutj (k_j) \to Z_\mu$ vertex (triangle diagram), where
the neutralino $i$ carries momentum $k_i$, does in general yield new
structures beside the ones present at tree-level. Therefore in general
the induced one-loop corrected $\tilde\chi^0_i \tilde\chi^0_j Z$ would have the
general form
\beqn
{\cal {L}}_{\neuti \neutj Z}^0 \to 
{\cal {L}}_{\neuti \neutj Z}^0 (g_Z \to g_Z^{{\rm eff}}, N \to N^{{\rm eff}}) + \neuti \left( K_1^{L,R} k^+_\mu + K_2^{L,R} k^-_\mu\right) P_{R,L} \neutj Z^\mu, \; \quad \; k^\pm= k_i \pm k_j.
\label{eq:k12lr}
\eeqn
$K_{1,2}^{L,R}$ are new coupling strengths. For on-shell $Z$, like in
$\nnzz$, the structure $k^+_\mu$ is of no relevance (as a consequence of the on-shell spin-1
condition), while for $i \neq j$ a structure with $k^-_\mu$ could be
present. One needs therefore to make sure that such new Lorentz
structures give a negligible contribution. While these new Lorentz
structures are necessarily ultraviolet finite without the need for
renormalisation, corrections to a tree-level structure need, in general,
renormalisation. An important ingredient is in fact given by the
counterterms to the different parameters entering the process and the
wave functions renormalisation, both effects calling for the evaluation
of some 2-point functions. This is an example of the universal
character of these contributions. In many instances these 2-point
functions are sufficient for the evaluation of the universal form
factor (as in $f \tilde{f} \neuto$) without reference to the nature
of the fermion/sfermion pair since the universal part only refers to the
$\neuto$. For $\neuti \neutj Z$, the 1PI vertex function is also
needed in order to yield a finite result. We have already encountered this situation
for the definition of the effective $\neuto \neuto Z$
coupling\cite{Boudjema:2011ig}.

\subsection{Renormalisation}

\noi In~\cite{baro07,Sloops-higgspaper,baro09,boudjema_chalons1} we
gave a detailed presentation of our procedure for the renormalisation
of all the sectors of the MSSM as implemented in our code for the
automatic evaluation of one-loop corrections,
\texttt{SloopS}~\cite{Sloops-higgspaper,baro09,Boudjema:2005hb}. We
stick to an on-shell scheme generalising what is done for the
Electroweak Standard Model~\cite{grace-1loop}. All fermion and gauge
boson masses are defined on-shell and the electric charge is defined
in the Thomson limit. With our input parameters for the masses of the
standard model fermions, the effective electric charge at the scale
$M_Z$ amounts to a correction of about $6.5\%$, therefore the running
of $\alpha$ alone in $\nnzz$ would give a correction of about $13\%$.
In the Higgs sector we take the mass of the neutral pseudo-scalar
Higgs $M_A$ as input while $\tb$ is defined, as usual\cite{Sloops-higgspaper}, from the decay
process $A^0 \to \tau \tau$. Other schemes for $\tb$ are possible~\cite{Sloops-higgspaper}. 
For $\nnzz$, in particular in
the higgsino case, the $\tb$ scheme dependence is very mild and we will
not discuss it here. In the case at hand what is most important in the renormalisation procedure are
the key parameters that enter the neutralino sector, namely $M_1, M_2,
\mu$. In \SloopS{} the default scheme is to choose two
charginos masses $\mchargopm$ and $\mchargtpm$ as input to define
$M_2$ and $\mu$ and one neutralino mass to fix $M_1$.  In this scheme,
the counterterms for the relevant parameters are~\cite{baro09}
\begin{eqnarray}
\delta M_{2}&=&\frac{1}{M_{2}^{2}-\mu^{2}}\Bigg(
\left(M_{2}\mchargo^2-\mu \det X\right)\frac{\delta\mchargo}{\mchargo}
+ \left(M_{2}\mchargt^2-\mu \det
  X\right)\frac{\delta\mchargt}{\mchargt} \nonumber\\
& &- M_{W}^2\left(M_{2}+\mu s_{2\beta}\right)\frac{\delta
M_{W}^2}{M_{W}^2} - \mu M_{W}^{2} s_{2 \beta}c_{2 \beta}
 \frac{\delta t_\beta}
{t_{\beta}}\Bigg),\nonumber\\
\delta \mu &=&\frac{1}{\mu^{2}-M_{2}^{2}}\Bigg(
\left(\mu\mchargo^2-M_{2} \det X\right)\frac{\delta\mchargo}{\mchargo}
+ \left(\mu\mchargt^2- M_{2}\det X\right)\frac{\delta\mchargt}{\mchargt} \nonumber\\
& &- M_{W}^2\left(\mu+M_{2} s_{2\beta}\right)\frac{\delta
M_{W}^2}{M_{W}^2} - M_{2} M_{W}^{2} s_{2 \beta} c_{2 \beta}
\frac{\delta t_\beta}
{t_{\beta}}\Bigg),\\
\delta M_{1}&=&\frac{1}{N_{1i}^{2}}\Big(\delta
m_{\tilde\chi_{i}^{0}}-N_{2i}^{2}\delta
M_{2}+2N_{3i}N_{4i}\delta\mu\nonumber
\\& &-2N_{1i}N_{3i}\delta
Y_{13} - 2N_{2i}N_{3i}\delta Y_{23} - 2N_{1i}N_{4i}\delta Y_{14} -
2N_{2i}N_{4i}\delta Y_{24}\Big)\, , \label{eq:deltam1}
\end{eqnarray}
with $\det X=\big(M_{2}\mu-M_{W}^{2}s_{2\beta}\big)$.
$\delta m_{\tilde{\chi}_{i}^{0}}$ is the counterterm of the $i$th
neutralino defined entirely from its self-energy.
In general, $\delta O$ represents the counterterm for the parameter $O$. Note that both
$\delta M_2$ and $\delta \mu$ could also be defined from the neutralino
sector, by a simple generalization of Eq.~\ref{eq:deltam1}.  The
definition of these counterterms reveals the presence of {\em denominators}
such as $\big(M_2^2-\mu^2\big)$ or $N_{ij}^2$. One should therefore
avoid such schemes in situations in which these denominators are very
small. For example, if the LSP has a very small bino component
one should in principle avoid taking its mass to define $M_1$ but rather choose a neutralino where
this component is not negligible. This being said, considering that
we are aiming primarily at finding a good approximation for the
higgsino case, the scheme dependence as concerns the best choice for
defining $M_1$ is not an issue, as we shall see. On-shell
renormalisation also requires that no-mixing between different physical 
fields remains after renormalisation of the
parameters and that the physical fields are such that the residue
at the pole of the propagators is one. This is achieved through wave 
function renormalisation for the neutralinos~\cite{baro09}
\begin{eqnarray}
\label{eq:wfrneut} \tilde{\chi}_{i}^{0} \to
\tilde{\chi}_{i}^{0}+\frac{1}{2}\sum_j \left(\delta
Z_{ij}P_{L}+\delta Z^{*}_{ij}P_{R}\right)\tilde{\chi}_{j}^{0}.
\label{eq:deltaZDeltaN}
\end{eqnarray}
These wave function renormalisation constants are particularly important for the definition of
the effective couplings.

\subsection{Implementation of the corrections for $\neuti \neutj Z$}
From what we have just seen, the form factor includes the effects due to
the renormalisation of the gauge couplings but also the effects due
to the mixing between fields at one-loop (including renormalisation of the
weak mixing angle and the mixing between the neutralinos). These are
implemented through the self-energy 2-point functions of the various
fields. To sum up, as advertised earlier, the effective form factor
vertex is obtained by substituting $g_Z \to g_{\neuti \neutj Z}^{{\rm eff}}$
and $N \to N + \Delta N$ with
\beqn
 \Delta N_{ij} &=&\frac{1}{2} \sum_{k} N_{kj} \delta Z_{ki}, \quad 
      (i,j,k)=1\dots 4, \nonumber \\     
 g_{\neuti \neutj Z}^{{\rm eff}}&=&g_Z \bigg(1+ \Delta g_Z + \Delta
g_{\neuti \neutj Z}^{\bigtriangleup}\bigg). 
\eeqn
\noi $\delta Z$ represents the various wave function renormalisations
for the neutralino system (see Eq.~\ref{eq:deltaZDeltaN}) obtained
solely through the set of two-point functions relative to the
self-energies of the neutralinos. As such all arguments of these
two-point functions are evaluated at the pole mass of the neutralinos.
The full expressions are given in~\cite{baro09}. \\
For the overall coupling $g_Z$ we see that it involves two parts,
$\Delta g_Z$ and $\Delta g_{\neuti \neutj
  Z}^{\bigtriangleup}$. Similarly to the shift $\Delta N$,
$\Delta g_Z$ is expressed solely in terms of the self-energies of the
neutral gauge bosons. We have
\begin{align}
  \Delta g_Z \equiv  \Delta g_Z(M_Z^2) =&\frac{1}{2}\left(\Pi'_{\gamma\gamma}(0)-2\frac{s_W}{c_W}
    \frac{\Pi_{\gamma Z}(0)}{M_Z^2}\right)+\frac{1}{2}
  \left(1-\frac{c_w^2}{s_w^2}\right)
  \left(\frac{\Pi_{ZZ}\left(M_Z^2\right)}{M_Z^2}-
    \frac{\Pi_{WW}\left(M_W^2\right)}{M_W^2}\right) \notag \\
  &-\frac{1}{2} \Pi^\prime_{ZZ}\left(M_Z^2\right),
  \label{eq:deltagz}
\end{align}
where $\Pi_{V V'}$ with $V,V'=W,Z,\gamma$ denotes the self-energies of
the gauge vector bosons. The combination of $\Delta g_Z$ and $\Delta N$,
defined solely from two point-functions, does not lead to a finite
result.  In order to get a finite result we need to add a genuine three-point function contribution which we have labeled $\Delta g_{\neuti \neutj Z}^{\bigtriangleup}$. We have decided to extract this
contribution from the amplitude of the one-loop transition
$\neutj \to \neuti Z$ by identifying the corrections to the coefficients 
of the Lorentz structures that are already present at the tree-level.\\
It is important to stress that, for the form factor
approximation, we only take into account leptons, quarks and their
superpartners circulating into the loops. Loops involving gauge bosons and their superpartners
have always been problematic and the problems are present even in the SM 
in the case of the $Z f \bar{f}$ process. In fact, in the approach we 
are using, it is difficult to extract a gauge independent value. This also leads   to problems with
unitarity. This would then require to include at least part of the box
contribution but this contribution cannot be described in the
simplified form factor approach. \\

The alert reader will have noticed that we have identified $\Delta
g_Z$ with $\Delta g_Z(M_Z^2)$. This is what would have been used for
the decay $Z \to f \bar f$ and, since the $Z$ is on-shell for $\nnzz$,
this is appropriate. Note that in~\cite{Boudjema:2011ig} we had improved on
this by using for $\neuto \neuto \to Z \to f \bar f$ a
running $\Delta g_Z$ evaluated at the invariant mass of the $f \bar f
$ system. Here, in the same vein we have implemented the $g_{\neuti
\neutj Z}^{{\rm eff}}$ for two different scales that enter the contribution $\Delta
g_{\neuti \neutj Z}^{\bigtriangleup}$. The scale relates to value of
the invariant mass of the would be, in $\nnzz$, intermediate
neutralino. As we will see when performing the calculations in
different scenarios, the difference in the corrected (effective)
calculation is small between the two choices of scales that we are
about to define.
\begin{itemize}
\item In the default implementation of the form factor approach , the
  $\neuto \neuti Z$ triangle vertex is evaluated on the mass shell for
  all three particles.  Although $\neuti \to \neuto Z$ (that is needed
  for $\nnzz$) is not always open kinematically, one may still, at the
  amplitude level, evaluate this transition with all external momenta
  on-shell. An advantage of this, is that once a model is defined and
  therefore all masses of neutralinos known, the form-factor is given
  once and for all for this model and could be applied to any kind of
  process where the vertex $\neuto \neuti Z$ is involved. In this
  implementation the effective coupling is simply denoted $g_{\neuti
    \neutj Z}^{{\rm eff}}$ and $Q^2_{\neuti}=m_{\tilde\chi^0_i}^2$.
\item One could think of slightly adapting the form factor to the
  kinematics of the $\nnzz$ process by taking into consideration that
  for $\nnzz$, the $\neuti$ in the vertex $\neuto \to \neuti Z$ is
  off-shell with invariant mass $Q^2_{\neuti} \sim M_Z^2-m_{\neuto}^2$
  (taking into account the small relative velocity of the LSP). In
  this case the effective coupling is denoted $g_{\neuti \neutj
    Z}^{{\rm eff}}(Q^2_{\neuti})$
\item In order to quantitatively compare the effect of different
  implementation of the vertex, we have also implemented the new
  Lorentz structure (the $k^-$ terms) together the effective coupling
  $g_{\neuti \neutj Z}^{{\rm eff}}(Q^2_{\neuti})$. We will refer to
  this approximation as $\Delta_{f \tilde{f}}$.  $\Delta_{f \tilde{f},
    no\;k^-}$ has no new Lorentz structure and corresponds therefore to
  $g_{\neuti \neutj Z}^{{\rm eff}}(Q^2_{\neuti})$.
\end{itemize}
As we will see, the effect of the induced new Lorentz structures is
totally negligible. This is a welcome feature since codes for the
calculations of tree-level cross sections based on a tree-level
Lagrangian can be used without implementing new structures and new
rules, we will only need to pass the modified overall effective
couplings. 
When we compare our results for the different approximations, FF
stands for the form factor approach implemented directly in a
tree-level calculator by exchanging the tree-level coupling by the
improved one. Modifying the coupling in a cross
section evaluator code from $g_Z$ to $g_Z (1+ \delta g_Z)$ will
inevitably incorporate contributions of order $\left(\delta
  g_Z\right)^2$. These should be small if the one-loop
contribution is perturbative and, thus, makes sense. Nonetheless for
corrections of order $10-20\%$ of the tree-level calculation a
form-factor implementation in a tree-level code that incorporates
${\cal O}( (\delta g_Z)^2)$ can be off by $2-4\%$. We should allow for
this when we compare the results with the full one-loop calculation
which does not include higher other terms.

Contrary to the FF, in the implementation  of $\Delta_{f \tilde{f}}$ and $\Delta_{f
  \tilde{f}, no\;k^-}$ no quadratic terms of type $(\delta g_Z)^2$ are present. Therefore,
beside the kinematics, with  $\Delta_{f \tilde{f}}$ and $\Delta_{f
  \tilde{f}, no\;k^-}$ we are following what
is implemented within a full one-loop calculation, save for the fact
that only leptons, quarks and their superpartners  are
kept in the loops.

Apart from the $\Delta_{f \tilde{f}}$ correction, the full one-loop
calculation includes
\begin{itemize}
\item the set of all 2-point, 3-point and 4-point function contributions not
  involving leptons, quarks and their superpartners. This set will be
  referred to as $(\Delta + \Box)_{no\; f\tilde{f}}$,
\item box diagrams involving leptons, quarks and their superpartners
  as depicted in the last diagram of
  Fig.~\ref{fig:1loop-diagrams},
\item one-loop corrections to the $s$-channel Higgs exchange
  contribution which is generally very small.
\end{itemize}

\section{Analysis at one-loop}
We now analyse the performance of the approximations compared to the
results of the full one-loop corrections for all four types of
scenarios that we described earlier when presenting the tree-level
cross sections.  The cross sections are evaluated as previously for $v \simeq 0.2$. When analyzing  the one-loop
results it is important to recall  the tree-level behaviour
and the LSP content as shown in
Fig.~\ref{fig:tree-level} to which we urge the reader to refer to alongside the loop corrections
we discuss below.

\subsection{Higgsino-like cases}
\begin{figure}[h]
\centering
\includegraphics{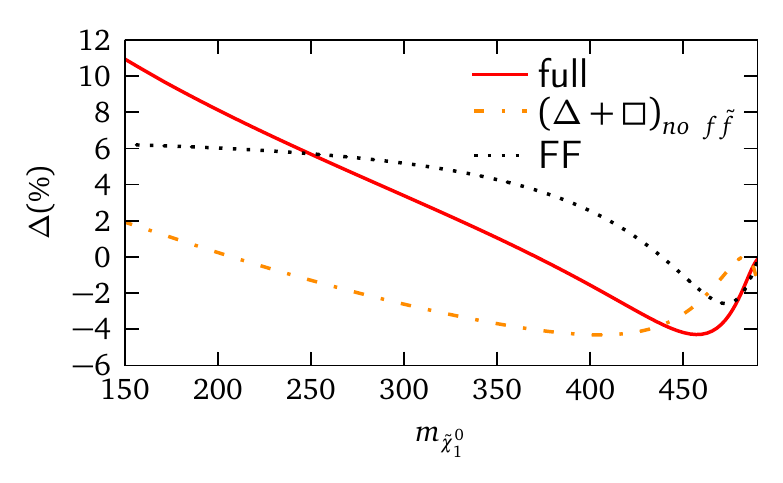}
\includegraphics{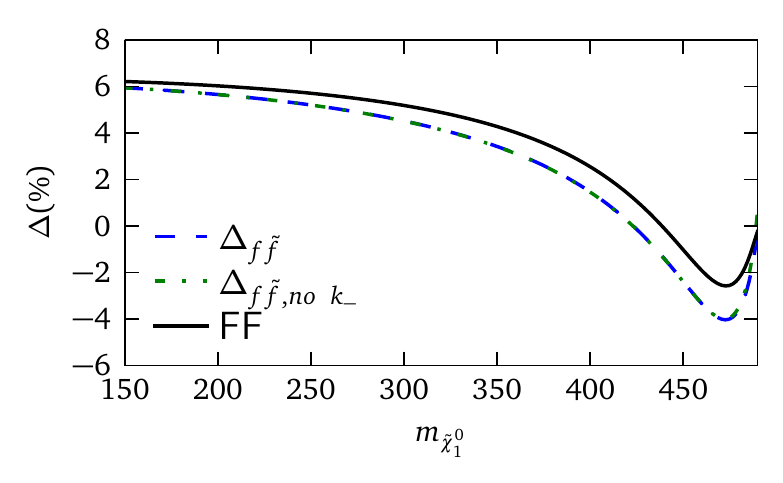}
\caption{{\em Radiative corrections as a function of the neutralino
    LSP mass for different approaches. Up to LSP masses around
    400~GeV, the LSP is dominantly higgsino. Around 450~GeV the LSP
    turns into a bino-like LSP. The panel on the right compares the
    different implementations of the form factor approach (see the
    text for the meaning of the labeling).}}
\label{fig:HinoToBino}
\end{figure}
\noi As we can see from Fig.~\ref{fig:HinoToBino}, in the case of a
higgsino-like LSP, the full one-loop radiative corrections range from 
11\% (for the lightest masses) to -3\% (for the heaviest LSP). The drop in the 
correction as the mass of the LSP
increases is smooth. As the nature of the LSP turns into bino-like, we see a
turn-over in the percentage correction. In the range of LSP mass between 450
and 500~GeV this correction increases slightly. This trend, with the turn over, is
reproduced with the form factor (FF) approximation. In fact, the result
that we obtain using the form factor is not far from the full
correction even if it is more ``flat'' for small (less than 300~GeV)
LSP masses. The largest discrepancy is observed for the lightest
masses, 150~GeV, where we have a difference slightly above $4\%$. Otherwise the
difference between the full result and the FF is well within $4\%$. A
naive implementation through a running of $\alpha$ would accidentally
be not a bad approximation for masses around 150~GeV, since this
amounts to about a $13\%$ correction.  This implementation would however be off as the
mass increases. Moreover, this will not show as much variation and structure as the
effective FF and the full one-loop correction suggests. Note that the
contribution of the
$W/\tilde\chi^\pm$ boxes are small, they are not larger than $4\%$. The $s$-channel
contribution (Higgs exchange) in this case is totally negligible. As for the
different implementations of the effective approach, we see that the addition of a
new Lorentz structure ($k^-$ terms) is totally negligible. For
most of the mass range, in particular for the whole range where the $\nnzz$
cross section is large and the LSP is dominantly higgsino, the form factor approach and the $\Delta_{f
\tilde f}$ agree very well. The largest difference shows up in the (uninteresting) bino-like region 
and amounts to no more than $1\%$. 

\begin{figure}[h]
\centering
\includegraphics{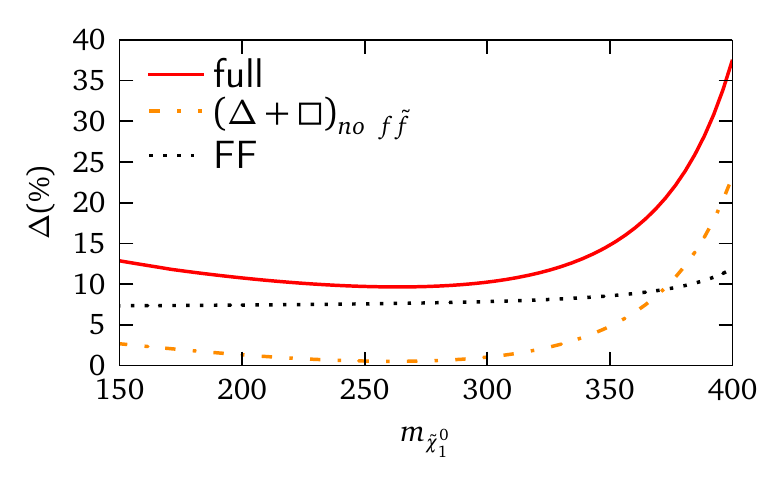}
\includegraphics{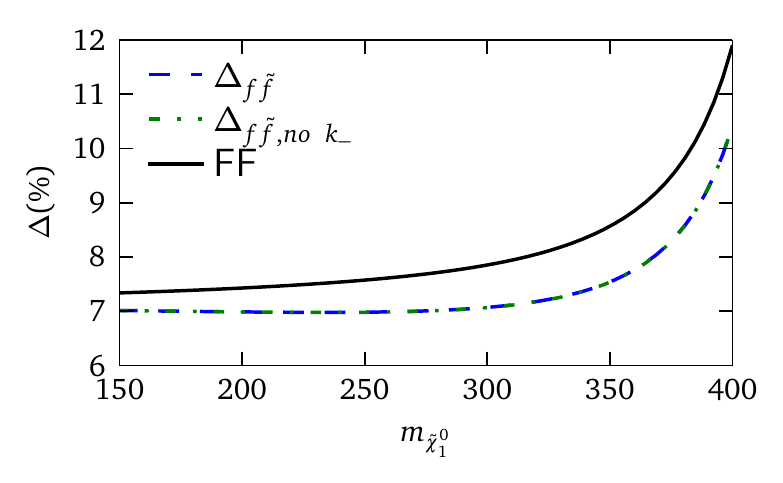}
\caption{{\em Higgsino LSP with some wino component, corresponding to
    the higgsino-like 2 scenario. The labeling is the same as in
    Fig.~\ref{fig:HinoToBino}. }}
\label{fig:HinoToWino}
\end{figure}

We now turn to the second scenario of higgsino-like LSP's and plot the
results in Fig.~\ref{fig:HinoToWino}. The difference with the previous
case is that the LSP picks up more and more wino component as the mass
the LSP increases.
Compared to the previous case, the  contamination due to the wino component 
starts much earlier in the sense that the higgsino fraction drops
more quickly. One must keep in mind that the tree-level cross section
$\nnzz$ is less than a tenth of $\nnww$ for LSP masses above 400~GeV. This is
the reason  we  only consider the corrections to the $\nnzz$
cross section for masses below 400~GeV.  Up to neutralino masses of
about 300~GeV the conclusions are the same as in the previous
case. For example in the range $\mneuto=200-300$~GeV, the difference
between the full correction and the form factor approach is less than
$4\%$. The difference increases very fast past 300~GeV. At 400~GeV,
the FF correction is about 12\% while the full correction is more than 35\%. On
the one hand, as the difference between the full one-loop and the FF starts growing, around 300~GeV, the ratio $\sigma_{\nnzz}/\sigma_{\nnww}$ gets
smaller, meaning that $\nnzz$ is not so relevant for a relic density calculation. On the other hand we
note that for the same reason the box contribution starts picking up
and is certainly not negligible in this region. As the
wino component increases, $\nnww$ becomes more and more important, consequently rescattering effects become important so that $\nnzz$ is
induced through $\nnww$ that rescatters to give $W^+W^- \to ZZ$. This is exactly the
contribution of the wino/chargino boxes (see third diagram in
Fig.~\ref{fig:1loop-diagrams}). We will see this more explicitly when
we will look at the loop corrections for a wino-like LSP annihilating
to $ZZ$.

As for the previous case, any one-loop induced new Lorentz
structure is totally negligible (see Fig.~\ref{fig:HinoToWino}). 
Once more, the difference between the
FF implementation (all particles on their mass-shell) and the vertex
insertion with the  invariant mass of the intermediate neutralino at
the correct kinematical value for $\nnzz$, is within 1\%. In fact, this difference
is much smaller than 1\% for the large $\nnzz$ cross sections
corresponding to the largest higgsino content.

\subsection{Bino-like LSP}
\begin{figure}[h]
\centering
\includegraphics{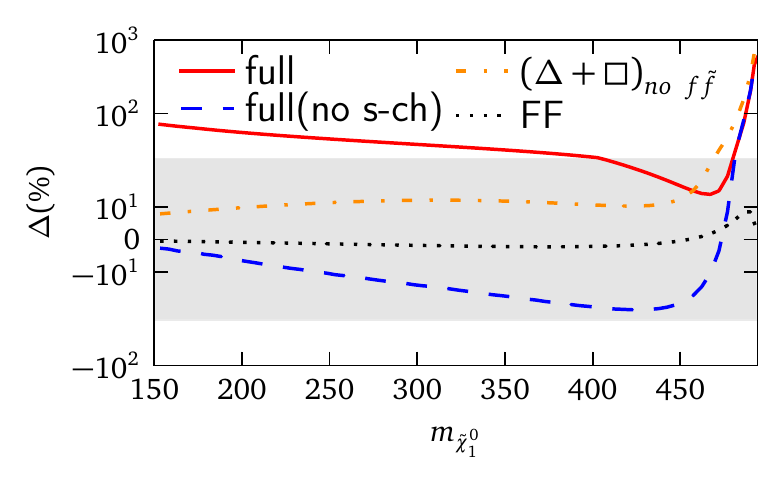}
\includegraphics{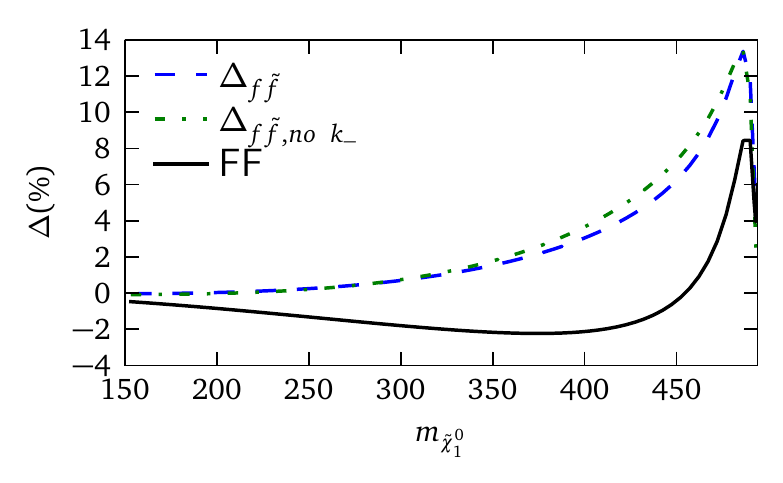}
\caption{{\em Results of a one-loop calculation in the case where the
    higgsino component is vanishing. For masses up to 450~GeV or so,
    the LSP is bino-like. Its nature changes suddenly to a wino-like
    LSP for masses greater than 450~GeV. Here we also show the results
    that one would obtain if the radiative corrections to the
    $s$-channel Higgs exchange are not taken into account, labeled  as     
    ``full (no s-ch)''. The
    rest of the labeling is the same as in
    Fig.~\ref{fig:HinoToBino}. The grey-shaded horizontal band corresponds to
    corrections within $\pm 20\%$, plotted using a linear scale. }}
\label{fig:BinoToWino}
\end{figure}
\noi Studying the radiative corrections to $\nnzz$ in such scenarios is not
particularly useful. In this scenario $f_H \sim 0$ and $\nnzz$
accounts for much less than $10\%$ of all annihilation channels. In
fact, up to masses around 450~GeV $f_B \sim 1$ and $\nnff$ dominates
the annihilation cross section. Even before the wino component fully
picks up, $\nnww$ increases (for masses past 450~GeV) and the
contribution of $\nnzz$ to the total annihilation rate is even
smaller. Nonetheless it is perhaps worth to see where the
corrections to $\nnzz$ stem from. Considering the smallness of the
tree-level $\nnzz$, it is more appropriate to indicate how $\nnzz$ is generated at
one-loop. Indeed, for most part of the parameter space in this
scenario the full correction is larger than 20\%, it even reaches
100\% for $\mneuto=150$~GeV and even more for the highest masses
when the LSP is wino-like. In the wino-like region (which we will
study more specifically in the next section), rescattering effects
through $\nnww$ become important, as we have seen in the last section. As Fig.~\ref{fig:BinoToWino} shows
for $\mneuto > 450$~GeV, the full correction is driven by the
$W$/$\tilde\chi^\pm$ loops, in particular by the box diagrams. For lower masses,
when the LSP is bino, this rescattering is relatively small. Yet the
FF fails completely in reproducing the full correction. What our study
shows is that it is important to correct the $s$-channel Higgs exchange
contribution in the bino case. Indeed leaving aside this correction gives a large discrepancy with the full correction. Note also that
the FF implementation marginally reproduces the non $s$-channel Higgs
exchange. One can therefore say that in this case the form factor
result is not very reliable both in the wino and in the bino
regions. Furthermore, our study reveals that there is an issue about
which invariant mass one implements for the intermediate neutralino
that is exchanged in the $t/u$ channels. Although this is smaller than
2\% for $\mneuto < 300$~GeV, the discrepancy increases to more than
4\%. This of course is small detail and tiny discrepancy compared to
the performance of the FF against  the full one-loop
correction. Once more, the effect of the new Lorentz structures in the
vertex are totally negligible.

\subsection{Wino-like LSP}
\begin{figure}[h]
\centering
\includegraphics{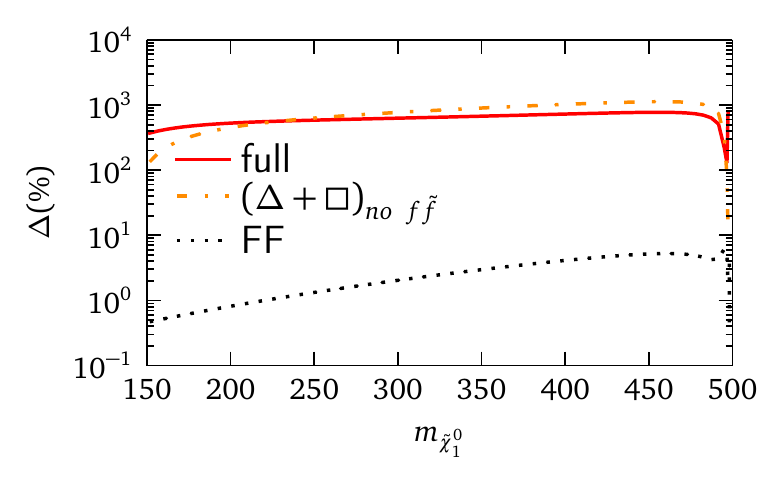}
\includegraphics{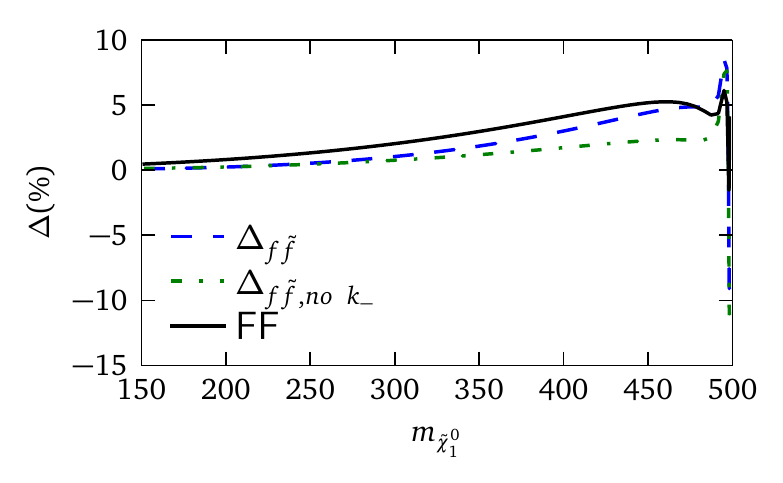}
\caption{{\em Same as in Fig.~\ref{fig:HinoToBino} for a wino-like LSP}  }
\label{fig:WinoToBino}
\end{figure}
In this scenario $\nnzz$ is extremely small
compared to the dominant $\nnww$ cross section, the two cross sections
are 6 orders of magnitude apart. Therefore for the relic density calculation 
there is no need to include $\nnzz$. The reason
we have looked at the one-loop corrections is just to make the
previous observations about the importance of rescattering $\nnww$
followed by $W^+ W^- \to ZZ$ more striking, as shown in Fig.~\ref{fig:WinoToBino}. We can see that for practically all masses the full
correction is driven by the $W$/$\tilde\chi^\pm$ loops and therefore the FF
approach based on the leptons, quarks and superpartners loops is totally negligible. Of
course in this case talking about radiative corrections does not make much sense, considering that the effect of the loops amounts to ``corrections'' in excess of a few hundred per-cent of the tree level cross section. As
we have argued previously, it is best to consider that for these cases
$\nnzz$ is induced through $\nnww$. The important message in this 
wino scenario is that one must perform a one-loop correction on
$\nnww$ since this is by far the largest cross
section. Co-annihilation processes should also be taken into account and we leave
these
studies for a forthcoming publication.
Fig.~\ref{fig:WinoToBino} also shows that there is little difference
between implementing the vertex correction through a full FF and a
$\Delta_{f \tilde{f}}$ and that, once more, the effect of a new
Lorentz structure although noticeable here, is below the 1\%. These
observation are of course an unimportant detail in view of the tiny
effect of the entire effective vertex correction.

\section{Renormalisation scheme dependence: the input neutralino masses}
The summary so far is that $\nnzz$ is an important annihilation channel as
long as one is in the higgsino region and that, in this region, it is important to have a
good prediction for this channel. In the higgsino-like scenarios we have seen that
the FF approximation is quite a good one. One should then address the
question of how much these conclusions, both for the full one-loop
calculation and the FF approximation, depend on the renormalisation
scheme. For the bino-like and wino-like scenarios this issue is of no
importance since $\nnzz$ is not an efficient annihilation channel.
\begin{figure}[htp]
\centering
\includegraphics{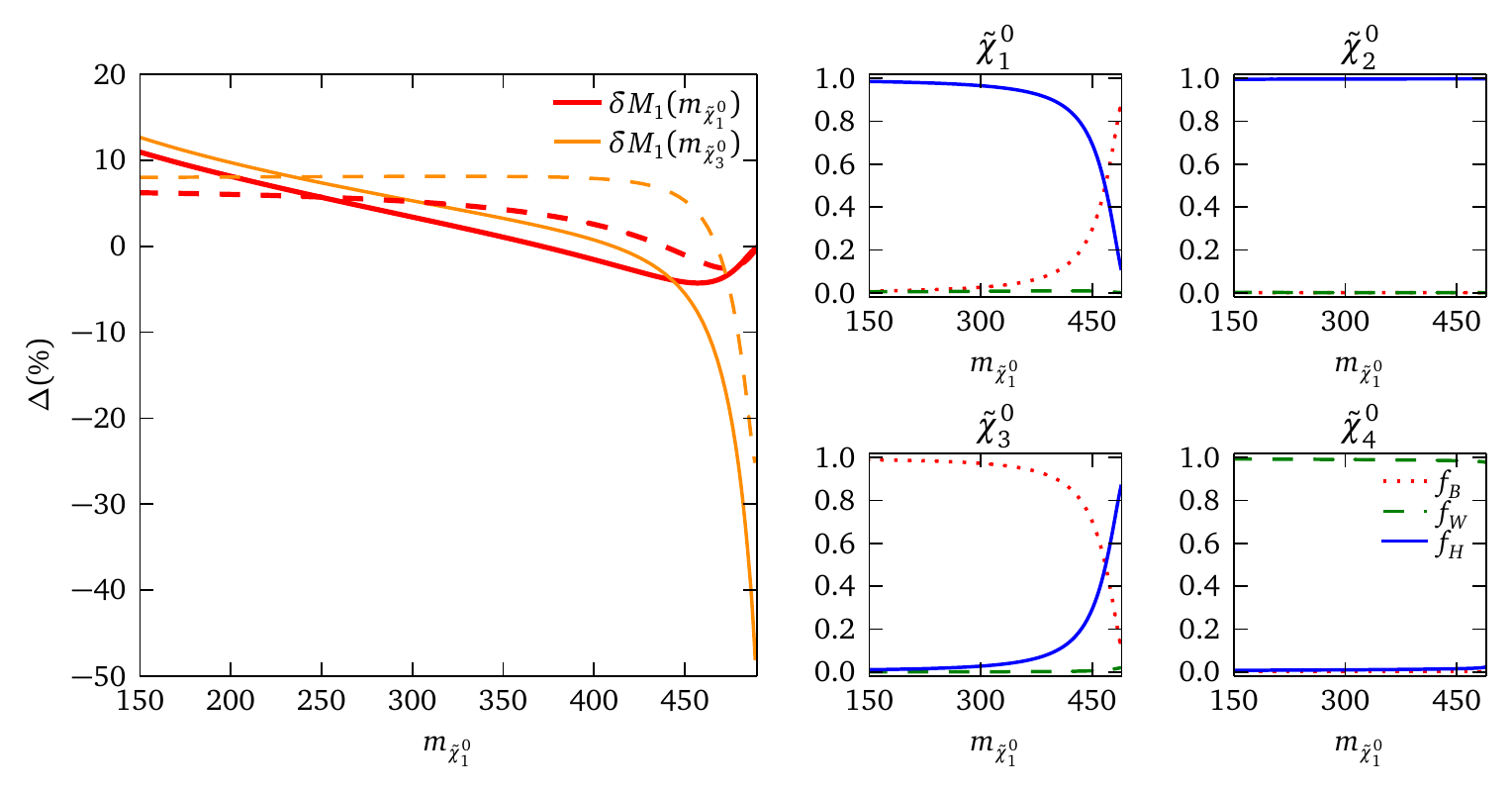}
\caption{{\em We compare the differences in the results for the
    percentage corrections between {\it i)} the default scheme where
    in the neutralino sector the LSP mass is taken as the input
    physical mass to reconstruct $M_1$ ($\delta M_1 (\mneuto)$) and
    {\it ii)} when the mass of the $\tilde\chi^0_3$ is taken as an input
    ($\delta m_{\tilde\chi^0_3}$).  We do this for both the full set of the
    radiative corrections (solid line) and the form factor approach
    (dashed lines). The small panels on the right indicate the
    different compositions for each of the four
    neutralinos. Throughout the range, $\tilde\chi^0_4$ is, for all
    purposes, a wino while $\tilde\chi^0_2$ is higgsino. Up to
    $\mneuto=450$~GeV,
    the LSP is higgsino-like and $\tilde\chi^0_3$ is bino-like, for
    masses beyond $450$~GeV the roles of $\tilde\chi^0_1$ and $\tilde\chi^0_3$ are
    swapped. }}
\label{fig:schemedep}
\end{figure}

\noi First we briefly review the $\tb$ scheme dependence. The default scheme is based on 
using $A^0 \to \tau \bar{\tau}$. Moving to a $\overline{{\rm DR}}$ scheme\cite{Sloops-higgspaper}
the changes are hardly noticeable. This is not surprising, recall our discussion on the $\tb$
dependence of the tree-level results.  We found that a change in
$\tb$ amounted to little effect on the cross section. In fact, the most
crucial scheme dependence concerns the choice of the neutralino mass to define the counterterm for $M_1$. In our default scheme, 
$\mneuto$ is used to reconstruct $M_1$,  
the bino parameter. In the higgsino limit, $f_H \sim 1$, this choice
does not, at first,  seem to be a good one since the bino component is very small. Indeed 
$N_{11} \sim 0$ in Eq.~\ref{eq:deltam1}. However, a one-loop calculation of $\nnzz$ is only crucial
in the higgsino limit where what matters is a good reconstruction of
$\mu$, or rather the higgsino component. This is quite nicely
extracted from the lightest chargino mass. 
Therefore in this limit since $\nnzz$ depends very mildly on $M_1$
there should be no difference between the different schemes that are used in the
neutralino sector to reconstruct $\delta M_1$. Fig.~\ref{fig:schemedep}
confirms these expectations in the scenario we call higgsino-like 1
where for masses up to $\mneuto=450$~GeV the LSP is dominantly
higgsino.  We find
that the $\delta M_1$ scheme dependence is totally negligible for the
full one-loop results up to $\mneuto=450$~GeV. If we compare  a scheme where $M_1$ is extracted from 
the most bino-like neutralino, $\tilde\chi^0_3$, with the default scheme, we find a difference that is within $1\%$ or 
so. The FF approximation follows the same trend although it looks like
the agreement between the full one-loop and FF is slightly better when
taking the LSP as input. Past a mass of $450$~GeV, the transition towards a more bino-like takes place
and the $\nnzz$ cross section decreases.   However once the transition occurs and the
amount of bino in the LSP becomes relevant, the way in which we
extract $M_1$ matters. We find that using $m_{\tilde\chi^0_3}$, instead of $\mneuto$ as input past
$450$~GeV leads to very large corrections and deviates significantly from the result of the default scheme. 
For $\mneuto \sim 480$~GeV the difference between using $\mneuto$ and $m_{\tilde\chi^0_3}$ for example is
about $20\%$. This is much larger than the difference between the full
one-loop and the effective coupling approach within the same scheme. The large $M_1$ scheme dependence in this 
bino configuration is directly related to the strong $M_1$ dependence of the tree-level cross section we pointed out in section~\ref{subsec:tree-level}. At the higgsino to bino transition point and above, it is perfectly
sensible to keep using $\mneuto$, the now bino-like LSP mass, to define $M_1$
($N_{11}$ in Eq.~\ref{eq:deltam1} is no longer small). We therefore suggest to always
use $\mneuto$ as the input parameter for $\nnzz$ no matter what the
composition of the LSP is. 
To summarize, whenever the higgsino component is large and $\nnzz$ is
of relevance, not only the FF is a good approximation but also, both for the 
full and the FF results, the scheme dependence is
negligible.

\section{Effects of very heavy sfermions and their non decoupling}
\begin{figure}[h]
\centering
\includegraphics[scale=1]{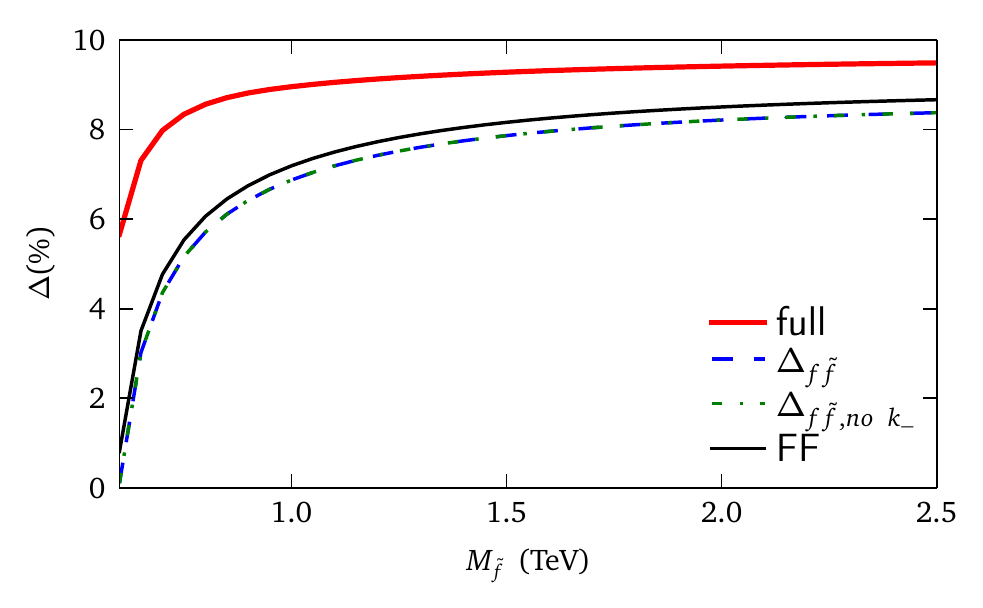}
\caption{{\em Dependence of the full one-loop and effective radiative corrections on
  the common sfermion mass for a higgsino-like LSP. These points corresponds to a neutralino of mass
  $m_{\tilde\chi_1^0} = 142.5$~GeV and components
  $(f_B,f_W,f_H)=(0.008,0.005,0.987)$.}}
\label{fig:sfMassDependence}
\end{figure}
We argued in the introduction to this paper and we discussed more at
length in~\cite{Boudjema:2011ig} when analyzing $\nnff$ that the
annihilation cross section of interest for the relic density exhibits
the effect of non decoupling of the very heavy sfermions which is a
consequence of supersymmetry breaking and the fact that $\mneuto \ll
m_{\tilde{f}}$. This non decoupling has been studied in a different
context
earlier\cite{feng_nondecoupling,randall_nondecoupling,nojiri_nondecoupling}. Fig.~\ref{fig:sfMassDependence}
shows how the correction increases as the mass of the common sfermion
mass increases from 500~GeV to 2.5~TeV. After a relatively
rapid rise, the increase is mild, almost leveling off for multi-TeV
masses of the sfermions.  The variation in the fermion/sfermion masses
is well reproduced by the effective couplings in the FF approach.

\section{Fixing the mass of the LSP}
Up to now we have presented our results as a function of the
neutralino LSP mass, for masses up to $450$~GeV. The LSP was for all
purposes in a ``pure'' state (higgsino, bino or wino). Past 450~GeV
its composition would, in most of the cases, change rather
drastically. It is interesting to also investigate the corrections and
the performance of the FF approximation by fixing the mass of the LSP
while varying its composition. We do this for three values of the LSP
mass, 110, 200, 400~GeV and generate points in the parameter space
that correspond to neutralino LSP masses within 2~GeV around these
values.
\begin{figure}[h]
\includegraphics{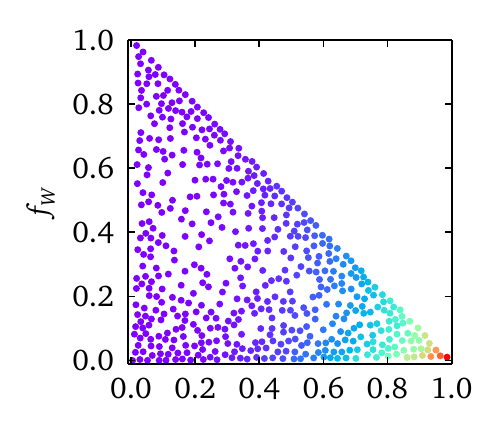}
\includegraphics{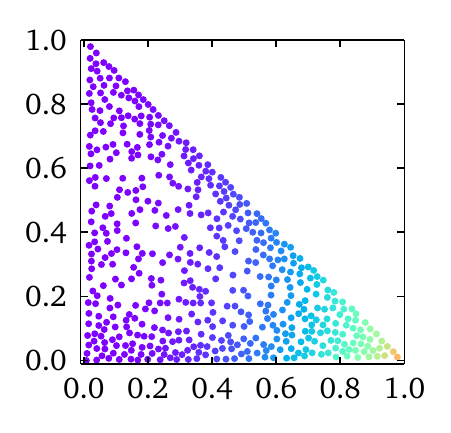}
\includegraphics{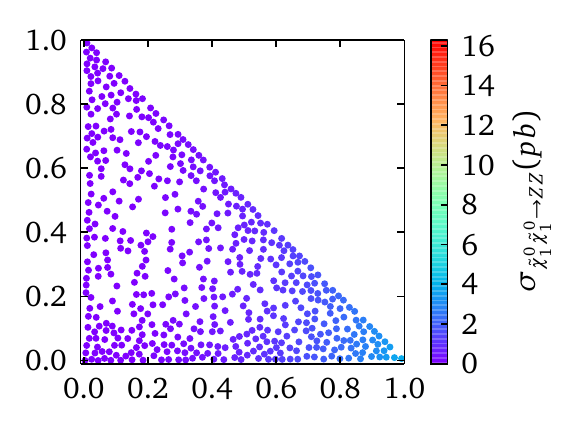}\\
\includegraphics{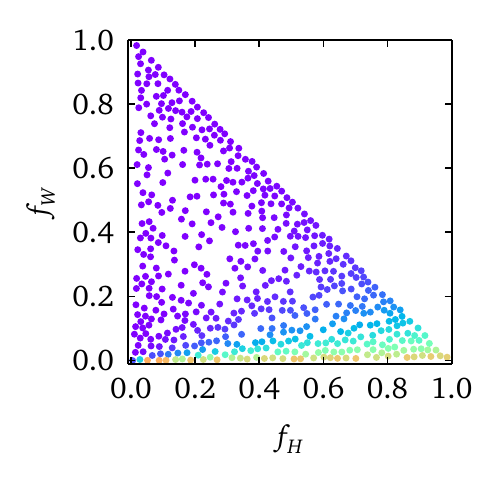}
\includegraphics{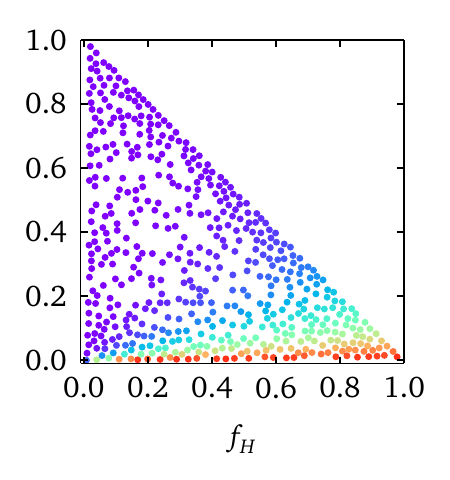}
\includegraphics{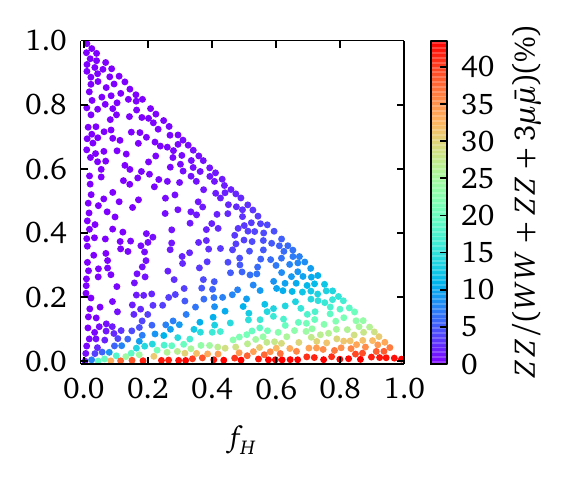}
\caption{{\em Annihilation cross section for the process
    $\tilde\chi_1^0\tilde\chi_1^0 \rightarrow ZZ$ according to its
    wino and higgsino content for a neutralino of mass (from left to
    right panel) 110 , 200 and 400 ($\pm 2 $)~GeV and a
    relative scattering velocity $v\simeq 0.2$. We plot the tree-level cross
    section for $\nnzz$ and its weight with respect to the
    $\tilde\chi_1^0\tilde\chi_1^0 \rightarrow W^\pm W^\mp$ and
    $\tilde\chi_1^0\tilde\chi_1^0 \rightarrow \mu\bar\mu$ cross
    sections.}}
\label{fig:fixedMassTree}
\end{figure}
  Fig.~\ref{fig:fixedMassTree} does confirm that the largest
cross sections do occur for the largest values of $f_H$. As the wino
component grows, the cross section decreases for all three values of
the LSP mass. In fact, as soon as $f_W>0.2$, the relative weight of the
$\nnzz$ (normalized to $\nnww$ and the annihilation into the three
leptons, without including possible co-annihilation processes) drops
below 10\%. We consider that below this relative weight it is not
important to get the full radiative correction since one should rather
concentrate on getting as a precise a result as possible for the
dominant annihilation process.

\begin{figure}[h]
\includegraphics{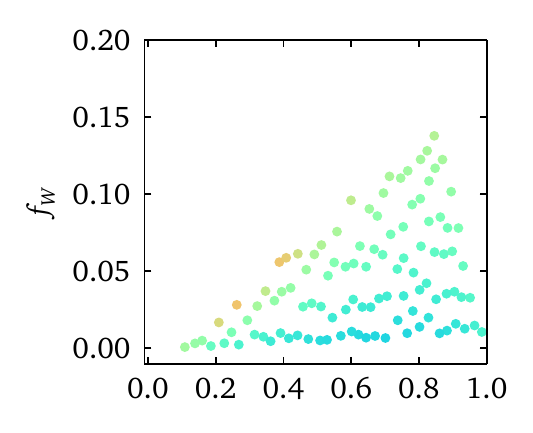}
\includegraphics{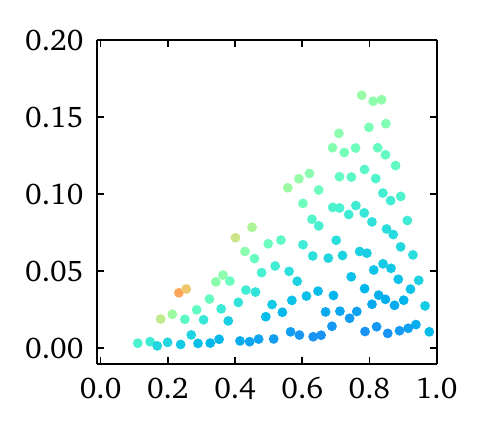}
\includegraphics{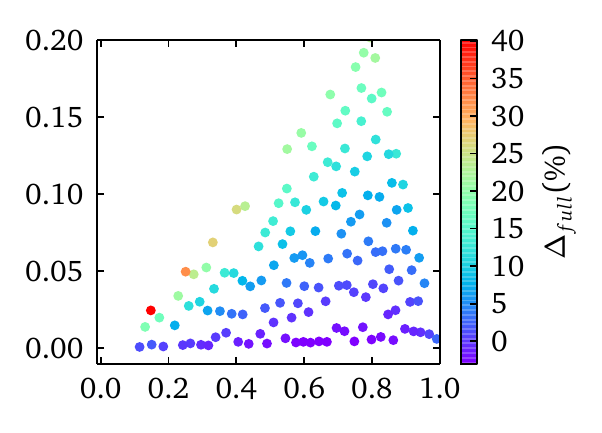}\\
\includegraphics{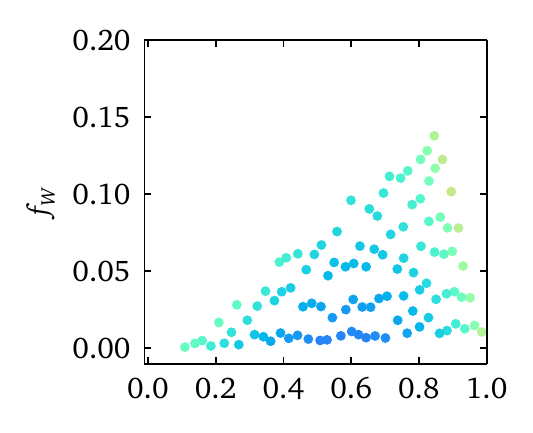}
\includegraphics{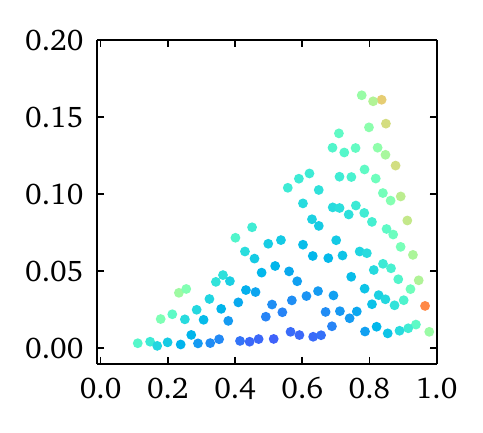}
\includegraphics{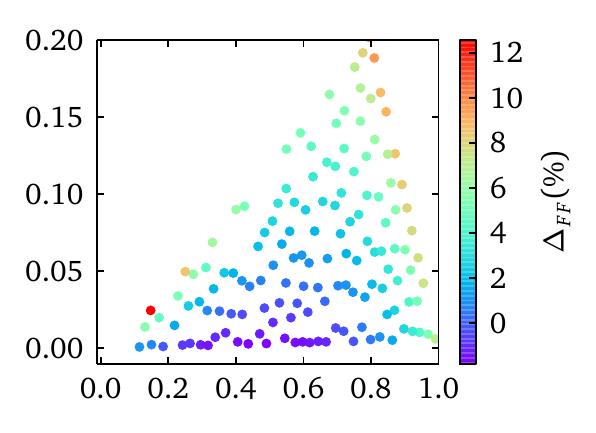}\\
\includegraphics{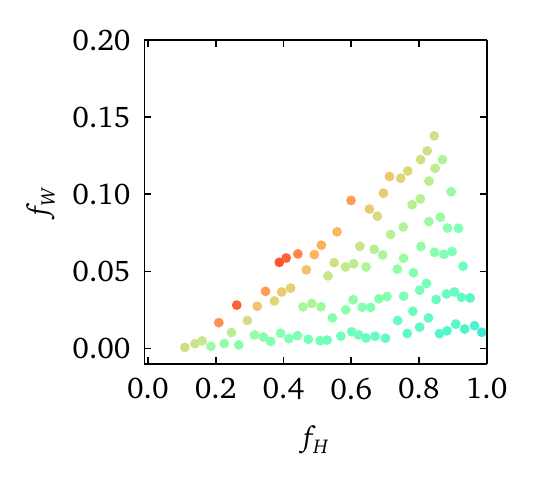}
\includegraphics{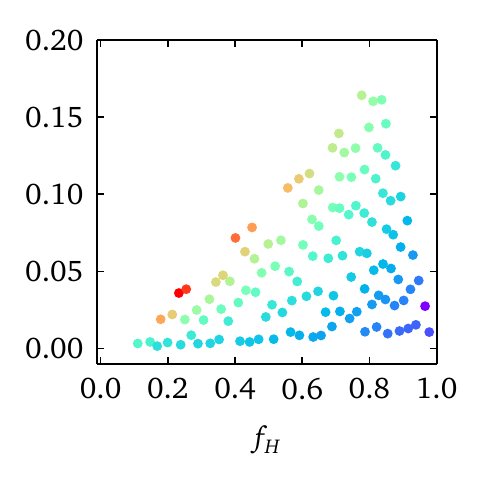}
\includegraphics{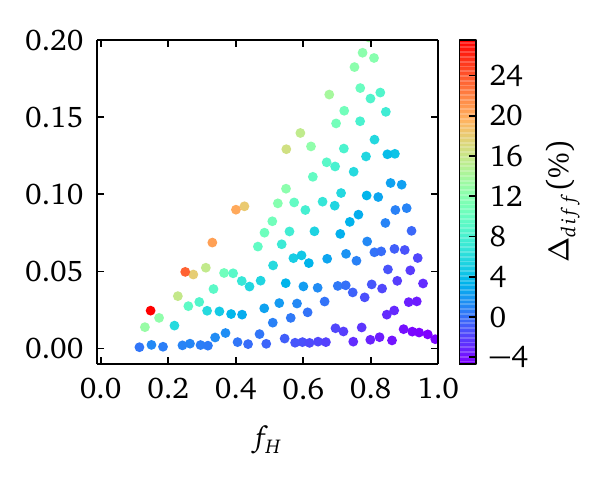}
\caption{{\em Relative one-loop contributions to the tree level annihilation
  cross section for the process $\tilde\chi_1^0\tilde\chi_1^0
  \rightarrow ZZ$ corresponding to the tree-level cross sections given in Fig.~\ref{fig:fixedMassTree}. The panels are ordered 
  in the same way as in  Fig.~\ref{fig:fixedMassTree}.  We plot the
  relative one-loop corrections obtained with the full one-loop
  calculation (first row), the form factor approach (second row) and the difference between the
  two (third row). We only consider those points for which the tree level cross
  section for the annihilation into $ZZ$ is at least the 10\% of the
  ``total'' cross section (see Fig.~\ref{fig:fixedMassTree}).}}
\label{fig:fixedMassLoop}
\end{figure}

We have then assessed, whenever $\nnzz$ is relevant, how the FF
performs and how large the
full radiative corrections can be. Fig.~\ref{fig:fixedMassLoop} shows
that the FF approach performs best in the higgsino corner. In that
corner the best results are obtained for $\mneuto=200, 400$~GeV where the
approximation is within $4\%$ of the full result. The worst agreement
is along a line with highest $f_W$ and lowest $f_H$, as
expected. However, with the restrictions we have imposed, namely  to consider
the radiative corrections only for points that have $\nnzz$ contribute
more than $10\%$, there are only very few points where the
disagreement is larger than 15\%.

\section{Summary}
Among all the constraints that are imposed on supersymmetry, for
example through scans on the parameter space, the value of the relic
density is the most stringent. The reason is most obvious. The current
value on the relic density as extracted from a combination of
cosmological measurements is at the per-cent level. Within the
standard thermal cosmological model, this very accurate measurement
translates into a very constraining bound on the cross sections
involved in the annihilation of the LSP dark matter into Standard
Model particles and hence on the underlying parameters of the
supersymmetric model.  Although this situation could be compared to
the impact that the precision LEP measurements had on constraining
many models of New Physics, the difference is that the experimental
precision on the relic density is not matched by  as precise
theoretical calculations. The analyses still use predictions on the
relic density based on tree-level calculations of the annihilation
cross sections and very often do not incorporate a theoretical
uncertainty that accounts for the missing higher order
calculations. Tools to perform one-loop calculations in supersymmetry
do exist and their exploitation for the relic density computation have
been achieved for many annihilation processes. It must however be
recognized that such full one-loop calculations are lengthy and bulky:
thousands of one-loop diagrams need to be evaluated, each one calling
large libraries for one-loop integrals.  Continuing our comparison
with LEP, it is important to inquire whether a large part of the full
one-loop corrections can be embedded in a minimal set of form factors
that correct the tree-level couplings. This set of improved couplings
can then be used for any process, not just the annihilation cross
sections but also, for example, decays of some particles. If this
programme can be realised it would be easy to exploit the same
existing tree-level codes that are used for the relic density
calculations. We initiated this program in a previous publication
where we focused our attention on a bino-like scenario for the LSP
where the
most important channel is $\nnff$. This has allowed to provide form
factors for two important couplings of the neutralinos: $\neuto f
\tilde{f}$ and $\neuto \neuto Z$. In this paper we have considered $\nnzz$ and in particular on extending the form factor library to
$\neuti \neuto Z$, where $\neuti$ can be any neutralino. The process $\nnzz$ is also what
characterizes a higgsino-like LSP. Of course,  annihilation into vector
bosons is more involved than annihilation into fermions and therefore
it is very important to compare the results of the form factor
approach to a full one-loop calculation. This is what we have
performed in this paper. The results we find are very encouraging. We
find that whenever annihilation to $ZZ$ is relevant for the relic
density, the form factor is a good approximation. In fact, for an almost 
pure higgsino the approach is very good. As the higgsino component 
degrades, the approximation becomes less and less reliable. This is
even more so when the contamination is due to a wino
component. However as the wino component gets large, $\nnzz$ becomes
inefficient as an annihilation channel. The cross section is tiny
compared to the then dominant $\nnww$ and therefore its weight in the
relic density calculation becomes more and more marginal as the form
factor approximation gets less and less precise. The dominance of
$\nnww$ is also the reason why the approximation fails. Indeed as we
have shown, in these wino scenarios, one-loop box diagrams that are
not accounted for by the form factor approach become important, if not
dominant. This is due to what we called the rescattering
effect. Indeed, in these cases $\nnzz$ is generated through $\nnww$
followed by $W^+ W^- \to ZZ$. The latter is a large cross
section. The leading contribution from such effects could most
probably be extracted in a compact form. We leave such improvement to
a forthcoming analysis. Although the one-loop calculation of $\nnzz$
is technically different from $\nnff$, the conclusions about the
performance of the form factors in the two cases are quite
similar. Whenever the cross sections are efficient channels that
contribute substantially to the relic density calculation, the form
factor approach is a good approximation, in the sense that we
reproduce the results within $4-5\%$. We have also checked that in these
cases there are no large theory uncertainty due to the renormalisation
scheme dependence. This good news encourages us to consider other
scenarios and provide more form factors. The next important step,
which we have already started, is a thorough investigation of the wino
case through the important annihilation channel $\nnww$. One should
also add to the list some important co-annihilation channels, in
particular those involving gauge bosons. The latter will be more
relevant for higher LSP masses in both the wino and higgsino cases. In
this paper we only considered LSP masses up to 500~GeV. The regime of
(multi) TeV LSP requires a different approach that must address the
large corrections of the Sudakov type and also in some cases the
Sommerfeld effect as discussed 
in~\cite{Ciafaloni:2013hya, Sommerfeldinclude,ArkaniHamed:2008qn,boudjema_chalons1}.

\vspace*{1cm}
\noi {\bf \large Acknowledgments}\\\\
We would like to thank Guillaume Chalons for many useful discussions.
This research was supported in part by the French ANR project
DMAstroLHC (ANR-12-BS05-0006), by the Labex ENIGMASS and by the
European Commission through the HiggsTools Initial Training Network
PITN-GA-2012-316704. AM would like to thank the ``Angelo della
Riccia'' Foundation for support.


\begin{thebibliography}{10}

\bibitem{Jarosik:2010iu}
N.~Jarosik {\it et al.}, Astrophys.\ J.\ Suppl.\ {\bf 192} (2011) 14
  [arXiv:1001.4744 [astro-ph.CO]].

\bibitem{planck_2013}
Planck Collaboration, P.~Ade {\em et~al.},
\newblock (2013), [arXiv:1303.5076 [astro-ph.CO]].

\bibitem{Percival:2009xn}
B.~A.~Reid {\it et al.} [SDSS Collaboration], Mon.\ Not.\ Roy.\ Astron.\ Soc.\
  {\bf 401} (2010) 2148 [arXiv:0907.1660 [astro-ph.CO]].

\bibitem{lhc_susy_limits_june2011}
see for example, S.~Caron for the ATLAS collaboration, arXiv:1106.1009
  [hep-ex].\\ J.~B.~G.~da Costa {\it et al.} [Atlas Collaboration], Phys.\
  Lett.\ B {\bf 701} (2011) 186 [arXiv:1102.5290 [hep-ex]]. \\ G.~Aad {\it et
  al.} [ATLAS Collaboration], Eur.\ Phys.\ J.\ C {\bf 71} (2011) 1682
  [arXiv:1103.6214 [hep-ex]].\\ S.~Chatrchyan {\it et al.} [ CMS Collaboration
  ],[arXiv:1107.1279 [hep-ex]].

\bibitem{LHCTeVScales}
For an updated summary giving limits from the LHC on masses of particles from
  different models of New Physics, in particular from searches of coloured
  objects, see P. Bargassa, {\em Strong SUSY production searches at LHC},
  Moriond 2013, Electroweak Session, {\tt
  https://indico.in2p3.fr/conferenceOtherViews.py?view=standard\&confId=9116}.

\bibitem{nonconventional-relic}
P.~Salati, Phys.\ Lett.\ B {\bf 571} (2003) 121 [arXiv:astro-ph/0207396].\\
  S.~Profumo and P.~Ullio, JCAP {\bf 0311} (2003) 006 [arXiv:hep-ph/0309220].\\
  F.~Rosati, Phys.\ Lett.\ B {\bf 570} (2003) 5 [arXiv:hep-ph/0302159].\\
  C.~Pallis, JCAP {\bf 0510} (2005) 015 [arXiv:hep-ph/0503080].\\ G.~B.~Gelmini
  and P.~Gondolo, Phys.\ Rev.\ {\bf D74} (2006) 023510
  [arXiv:hep-ph/0602230].\\ D.~J.~H.~Chung, L.~L.~Everett, K.~Kong and
  K.~T.~Matchev, arXiv:0706.2375 [hep-ph].\\ M.~Drees, H.~Iminniyaz and
  M.~Kakizaki, \textit{Phys.Rev.} {\bf D76} (2007) 103524,
  [arXiv:0704.1590[hep-ph]].\\ A.~Arbey and F.~Mahmoudi, JHEP {\bf 1005} (2010)
  051 [arXiv:0906.0368 [hep-ph]].

\bibitem{micromegas}
G.~B\'{e}langer, F.~Boudjema, A.~Pukhov, A.~Semenov, \textit{Comput. Phys.
  Commun.} {\bf 149} (2002) 103, hep-ph/0112278;\\ G.~B\'elanger, F.~Boudjema,
  A.~Pukhov, A.~Semenov, \textit{Comput. Phys. Commun.} {\bf 174} (2006) 577,
  hep-ph/0405253;\\ G.~B\'elanger, F.~Boudjema, A.~Pukhov, A.~Semenov,
  \textit{Comput. Phys. Commun.} {\bf 176} (2007) 367, hep-ph/0607059;\\
  G.~B\'elanger, F.~Boudjema, A.~Pukhov and A.~Semenov, Comput.\ Phys.\
  Commun.\ {\bf 177} (2007) 894.\\ G.~B\'elanger, F.~Boudjema, A.~Pukhov and
  A.~Semenov, Comput.\ Phys.\ Commun.\ {\bf 180} (2009) 747 [arXiv:0803.2360
  [hep-ph]].\\ G.~B\'elanger, F.~Boudjema, A.~Pukhov, A.~Semenov, \textit{
  Comput. Phys. Commun.} {\bf 185} (2014) 960 [arXiv:1305.0237 [hep-ph]]. \\
  {\tt http://lapth.cnrs.fr/micromegas}.

\bibitem{darksusy}
{\tt DarkSUSY}: P.~Gondolo \textit{et al.}, \textit{JCAP} {\bf 0407} (2004)
  008, astro-ph/0406204;\\ {\tt http://www.physto.se/$\sim$edsjo/darksusy/}.

\bibitem{superiso-relic}
{\tt SuperIso Relic}: A.~Arbey, F.~Mahmoudi, A.~Arbey and F.~Mahmoudi,
  1277 [arXiv:0906.0369 [hep-ph]]. \\ Comput.\ Phys.\ Commun.\ {\bf 182}, 1582
  (2011).\\ {\tt http://superiso.in2p3.fr/relic/}.

\bibitem{boudjema_gondolo}
F.~Boudjema, J.~Edsjo and P.~Gondolo, in {\em Particle dark matter} 325-344,
  Oxford University Press (2010) G.~Bertone, editor; 
  Matter and at the Colliders,'' [arXiv:1003.4748 [hep-ph]].

\bibitem{Hisano:2002fk}
J.~Hisano, S.~Matsumoto, and M.~M. Nojiri,
\newblock Phys.Rev. {\bf D67}, 075014 (2003), [arXiv:hep-ph/0212022].

\bibitem{Iengo:2009ni}
R.~Iengo,
\newblock JHEP {\bf 0905}, 024 (2009), [arXiv:0902.0688 [hep-ph]].

\bibitem{ArkaniHamed:2008qn}
N.~Arkani-Hamed, D.~P. Finkbeiner, T.~R. Slatyer, and N.~Weiner,
\newblock Phys.Rev. {\bf D79}, 015014 (2009), [arXiv:0810.0713 [hep-ph]].

\bibitem{Hryczuk:2011vi}
A.~Hryczuk and R.~Iengo,
\newblock JHEP {\bf 1201}, 163 (2012), [arXiv:1111.2916 [hep-ph]].

\bibitem{Sommerfeldinclude}
For a recent review see, A.~Hryczuk, Phys.\ Lett.\ B {\bf 699} (2011) 271
  [arXiv:1102.4295 [hep-ph]].

\bibitem{boudjema_chalons1}
N.~Baro, F.~Boudjema, G.~Chalons and S.~Hao, Phys.\ Rev.\ D {\bf 81} (2010)
  015005 [arXiv:0910.3293 [hep-ph]]. 

\bibitem{Ciafaloni:2013hya}
P.~Ciafaloni {\em et~al.},
\newblock JCAP {\bf 1310}, 031 (2013), [arXiv:1305.6391 [hep-ph]].

\bibitem{baro07}
N.~Baro, F.~Boudjema, A.~Semenov, \textit{Phys. Lett.} {\bf B660} (2008) 550,
  arXiv:0710.1821 [hep-ph].

\bibitem{baro09}
N.~Baro, F.~Boudjema, \textit{Phys. Rev} {\bf D80} (2009) 076010,
  [arXiv:0906.1665 [hep-ph]].

\bibitem{Freitas-relic-qcd}
A.~Freitas, Phys.\ Lett.\ B {\bf 652} (2007) 280 [arXiv:0705.4027 [hep-ph]].

\bibitem{HerrmannQCD}
B.~Herrmann and M.~Klasen, 
  in the Higgs Funnel,'' Phys.\ Rev.\ D {\bf 76} (2007) 117704 [arXiv:0709.0043
  [hep-ph]].\\ B.~Herrmann, M.~Klasen and K.~Kovarik, 
  Annihilation into Massive Quarks with SUSY-QCD Corrections,'' Phys.\ Rev.\ D
  {\bf 79} (2009) 061701 [arXiv:0901.0481 [hep-ph]].\\ B.~Herrmann, M.~Klasen
  and K.~Kovarik, 
  beyond scalar or gaugino mass unification,'' Phys.\ Rev.\ D {\bf 80} (2009)
  085025 [arXiv:0907.0030 [hep-ph]].

\bibitem{Boudjema:2005hb}
F.~Boudjema, A.~Semenov, and D.~Temes,
\newblock Phys.Rev. {\bf D72}, 055024 (2005), [arXiv:hep-ph/0507127].

\bibitem{Sloops-higgspaper}
N.~Baro, F.~Boudjema, A.~Semenov, \textit{Phys. Rev.} {\bf D78} (2008) 115003,
  arXiv:0807.4668 [hep-ph].

\bibitem{lanhep}
A. Semenov. {\it LanHEP --- a package for automatic generation of Feynman
  rules. User's manual.}; hep-ph/9608488. \\ A.~Semenov, \textit{Nucl. Inst.
  Meth. and Inst.} {\bf A393} (1997) 293;\\ A.~Semenov, \textit{Comp. Phys.
  Commun.} {\bf 115} (1998) 124;\\ A.~Semenov, hep-ph/0208011; \\ A.~Semenov,
  {\em Comput. Phys. Commun.} {\bf 180} (2009) 431, arXiv:0805.0555 [hep-ph].

\bibitem{feynarts}
J.~K\"ublbeck, M.~B\"ohm, A.~Denner, \textit{Comp. Phys. Commun.} {\bf 60}
  (1990) 165;\\ H.~Eck, J.~K\"ublbeck, \textit{Guide to FeynArts~1.0},
  W\"urzburg, 1991;\\ H.~Eck, \textit{Guide to FeynArts~2.0}, W\"urzburg,
  1995;\\ T.~Hahn, \textit{Comp. Phys. Commun.} {\bf 140} (2001) 418,
  hep-ph/0012260.

\bibitem{formcalc}
T.~Hahn, M.~Perez-Victoria, \textit{Comp. Phys. Commun.} {\bf 118} (1999) 153,
  hep-ph/9807565;\\ T.~Hahn, hep-ph/0406288; hep-ph/0506201.

\bibitem{looptools}
T. Hahn, {\tt LoopTools}, \verb+http://www.feynarts.de/looptools/+.

\bibitem{Boudjema:2011ig}
F.~Boudjema, G.~Drieu La~Rochelle, and S.~Kulkarni,
\newblock Phys.Rev. {\bf D84}, 116001 (2011), [arXiv:1108.4291 [hep-ph]].

\bibitem{Hollik_susyeff}
J.~Guasch, W.~Hollik and J.~Sola, JHEP {\bf 0210}, 040 (2002)
  [arXiv:hep-ph/0207364]. 

\bibitem{nojiri_nondecoupling}
S.~Kiyoura, M.~M.~Nojiri, D.~M.~Pierce and Y.~Yamada, Phys.\ Rev.\ D {\bf 58},
  075002 (1998) [arXiv:hep-ph/9803210].

\bibitem{Akcay:2012db}
A.~Chatterjee, M.~Drees, and S.~Kulkarni,
\newblock Phys.Rev. {\bf D86}, 105025 (2012), [arXiv:1209.2328 [hep-ph]].

\bibitem{CMS:2013dea}
CMS Collaboration, CMS-PAS-SUS-13-006 (2013).

\bibitem{TheATLAScollaboration:2013hha}
ATLAS Collaboration, ATLAS-CONF-2013-049, ATLAS-COM-CONF-2013-050 (2013).

\bibitem{ATLAS:2013rla}
ATLAS Collaboration, ATLAS-CONF-2013-035, ATLAS-COM-CONF-2013-042 (2013).

\bibitem{CMS:2013bda}
CMS Collaboration, CMS-PAS-SUS-12-022 (2013).

\bibitem{Kraml:2013mwa}
S.~Kraml {\em et~al.},
\newblock (2013), [arXiv:1312.4175 [hep-ph]].

\bibitem{Aad:2014vma}
G.~Aad {\it et al.} [ATLAS Collaboration], [arXiv:1403.5294 [hep-ex]],
  [arXiv:1402.7029 [hep-ex]].

\bibitem{grace-1loop}
G.~B\'{e}langer, F.~Boudjema, J.~Fujimoto, T.~Ishikawa, T.~Kaneko, K.~Kato,
  Y.~Shimizu, \textit{Phys. Rep.} {\bf 430} (2006) 117, hep-ph/0308080.

\bibitem{feng_nondecoupling}
H.~C.~Cheng, J.~L.~Feng and N.~Polonsky, Phys.\ Rev.\ D {\bf 56} (1997) 6875,
  [arXiv:hep-ph/9706438]; idem Phys.\ Rev.\ D {\bf 57} (1998) 152,
  [arXiv:hep-ph/9706476]. 

\bibitem{randall_nondecoupling}
E.~Katz, L.~Randall and S.~f.~Su, Nucl.\ Phys.\ B {\bf 536} (1998) 3
  [arXiv:hep-ph/9801416].

\end{thebibliography}

\end{document}